\documentclass[iicol,pdflatex,sn-basic,]{sn-jnl}

\usepackage{graphicx}%
\usepackage{multirow}%
\usepackage{amsmath,amssymb,amsfonts}%
\usepackage{amsthm}%
\usepackage{mathrsfs}%
\usepackage[title]{appendix}%
\usepackage{xcolor}%
\usepackage{textcomp}%
\usepackage{manyfoot}%
\usepackage{booktabs}%
\usepackage{algorithm}%
\usepackage{algorithmicx}%
\usepackage{algpseudocode}%
\usepackage{listings}%
\usepackage{subcaption}

\theoremstyle{thmstyleone}%
\theoremstyle{thmstyletwo}%

\theoremstyle{thmstylethree}%

\begin{document}

\title[Article Title]{Supervised tax compliance and evasion from a spatial evolutionary game perspective}


\author[1]{\sur{Qin Li}}

\author[2]{\sur{Ting Ling}}

\author*[2]{\sur{Minyu Feng}}\email{myfeng@swu.edu.cn}

\author[3]{Attila Szolnoki}

%\author[1,2]{\fnm{Third} %\sur{Author}}\email{iiiauthor@gmail.com}

\affil[1]{\orgdiv{Business College}, \orgname{Southwest University}, \orgaddress{\city{Chongqing}, \postcode{402460}, \country{China}}}

\affil[2]{\orgdiv{College of Artificial Intelligence}, \orgname{Southwest University}, \orgaddress{\city{Chongqing}, \postcode{400715}, \country{China}}}

%\affiliation[3]{Institute of Technical Physics and Materials Science, Centre for Energy Research, P.O. Box 49, H-1525 Budapest, Hungary.}

\affil[3]{\orgdiv{Institute of Technical Physics and Materials Science}, \orgname{Centre for Energy Research}, \orgaddress{\street{H-1525}, \city{Budapest}, \postcode{P.O. Box 49}, \country{Hungary}}}

%\affil[3]{\orgdiv{Department}, \orgname{Organization}, \orgaddress{\street{Street}, \city{City}, \postcode{610101}, \state{State}, \country{Country}}}


\abstract{Taxation constitutes a fundamental component of modern national economic systems, exerting profound impacts on both societal functioning and governmental operations. In this paper, we employ an interdependent network approach to model the coevolution between citizens and regulators within a taxation system that fundamentally constitutes a public goods game framework with complex interactive dynamics. In a game layer, citizens engage in public goods games, facing the social dilemma of tax compliance (cooperation) versus evasion (defection). Tax compliance supports the sustainability of public finances while tax evasion presents markedly stronger short-term incentives. In a regulatory layer, fair regulators punish tax evaders, while corrupt regulators keep silent due to bribes. Governmental regulatory interventions introduce critical institutional constraints that alter the traditional equilibrium of the game. Importantly, there exists a strategy update not only among citizens but also among regulators. Our results indicate that strengthening penalties can effectively curb tax evasion, and the influence of bribery on both tax compliance rates and the proportion of fair regulators is nonlinear. Additionally, increasing regulators' salaries and intensifying the crackdown on corrupt regulators can foster the emergence of fair regulators, thereby reducing tax evasion among citizens. The results offer practical policy implications, suggesting that balanced deterrence and institutional fairness are essential to sustaining compliance, and point to the need for future empirical validation and model extensions.}

\keywords{Tax Compliance, Corrupt Regulation, Spatial 
%Evolutionary Games, 
Populations,
Public Goods Games, Co-evolution}



\maketitle

\section{Introduction}\label{sec1}

Taxation, as a crucial pillar of national governance, traces its origins to tribute systems from ancient civilizations and has evolved alongside the development of state structures~\citep{1,2}. In modern society, taxation serves as the core source of public finance, providing the material basis for government services and acting as an important tool for regulating social equity. However, the conflict between tax compliance and tax evasion remains a fundamental dilemma faced by contemporary tax systems~\citep{3,4,5,mebratu2024theoretical}. In this study, tax evasion refers to the deliberate and illegal concealment of taxable income or falsification of information to reduce tax liability, which is distinct from tax planning -- the lawful arrangement of financial affairs to minimize tax burden within the boundaries of legal provisions. Following the definitions in recent accounting and compliance studies~\citep{kollruss2024hybrid}, tax evasion thus denotes a non-compliant behavior violating statutory obligations, while tax planning represents a legitimate optimization within the tax code. From the perspective of economic rational choice theory, this contradiction arises from the non-excludable and non-rivalrous nature of public goods: while paying taxes can maximize collective welfare, individual taxpayers may still choose to evade taxes in pursuit of short-term gains~\citep{6}.

Spatial evolutionary games, which combine network science with evolutionary game theory, provide an effective framework to understand the dynamic relationship between tax compliance and tax evasion behaviors in realistic models~\citep{7,8}. This approach has been shown to be a powerful tool for analyzing the emergence and persistence of self-organizing behaviors in natural and social systems, drawing interest from diverse disciplines, such as mathematical biology~\citep{nowak1992evolutionary,ohtsuki2006simple}, statistical physics~\citep{perc2017statistical, feng2024information}, mathematics~\citep{sun2023state,pi2024memory}, computational social science~\citep{tembine2009evolutionary, zeng2025complex, pi2025dynamic} and other disciplines~\citep{zhang2024impact,scata2016combining,ling2025supervised}. Importantly, in 2006, Nowak named five leading mechanisms that promote cooperation: kin selection, direct reciprocity, indirect reciprocity, group selection and network reciprocity~\citep{fiverules}. These fundamental mechanisms, along with subsequently discovered mechanisms, such as memory effects~\citep{wang2006memory,danku2019,xu2019}, multi-gaming~\citep{Szolnoki2014coevolutionary,dudong2016}, reputation~\citep{fu2008reputation,feng2023,xia2023}, rewards~\citep{Szolnoki2010reward,yan2024} and punishment~\citep{boyd2003,Szolnoki2011competition,zhang2025punishment}, collectively provide a solid theoretical foundation for understanding the evolution of cooperation in tax compliance behavior. Among all of these mechanisms, punishment is one of the most intensively studied mechanisms. In previous studies, the way of punishment was mainly characterized as individual (peer) or institutional (pool) punishment. Notably, other punishment mechanisms have also been explored, such as self-organized punishment~\citep{perc2012self,hua2023}, punishment with tolerance~\citep{Szolnoki2015tolerance}, inequity-averse punishment~\citep{wang2022,ding2025}, or mercenary punishment~\citep{lee2022}. Regardless of the form of punishment, the sanctions imposed by punishers represent a unique ability and, thus, a privilege. Nevertheless, this privilege carries an inherent risk of corrupt exploitation by enforcers.

Generally, citizens pay taxes to the government, which acts as a third party to punish tax evaders and maintain social stability. However, in reality, not all members of such institutions exercise their authority impartially, as corruption still exists within societal systems~\citep{zhang2023corruption}. The United Nations Convention Against Corruption states that corruption has become a serious social problem that threatens stability and security~\citep{un2004united}. It undermines democratic institutions and values, moral standards and justice, constituting a threat to a country's political stability and sustainable development. Recently, some researchers have focused on this problem. Verma and Sengupta created both deterministic and stochastic evolutionary game theory models to study bribery, discovering that asymmetric punishment scenarios can mitigate corruption under specific conditions~\citep{Verma2015}. Shi {\it et~al.} proposed a model of referee intervention involving corruption to explore its impact on punishment mechanisms. The results indicated that referee intervention always improves social efficiency, even within a completely corrupt system~\citep{Shi2022}. Furthermore, control mechanisms, such as mutual supervision between players and judges~\citep{shi2023bidirectional}, anti-bribery economic sanctions~\citep{huang2018evolution}, social ostracism~\citep{liu2021evolutionary}, and the strategic utilization of leaders' punitive authority and economic leverage~\citep{liu2022effects} have been proposed to mitigate the detrimental effects of corruption.

In recent years, several public goods game (PGG) models incorporating taxation mechanisms have been developed~\citep{shen2023evolutionary,griffin2017cyclic,wang2021tax}. However, these models often overlooked the critical role of corrupt regulation. Building on this foundation, this paper aims to answer the following questions:

\begin{enumerate}[1. ]

\item What is the effect of corrupt regulators on the evolution of tax compliance behaviors?

\item How do penalties, bribery ratio, regulatory fees, and the costs of corruption impact tax compliance rates and fair regulator's density?

\item What strategies can be employed in institutional design to optimize tax governance?

\end{enumerate}

To address these issues, we utilize the framework of evolutionary game theory to explore the evolution of citizen tax behavior under government supervision. Recent interdisciplinary studies published in Humanities and Social Sciences Communications have demonstrated how computational and evolutionary modeling can provide valuable insights into complex social behaviors and policy dilemmas,  including environmental cooperation, digital interaction, and corporate ethics~\citep{harring2021social, wu2025multiplayer, zeng2025authentic}. Since the game between citizens and regulators does not occur within the same network layer, we employ a multilayer network for modeling~\citep{wang2015,xiong2024,li2019competition}, by considering both citizens and government officials as nodes in separate networks. It comprises a game layer formed by citizens engaged in social dilemmas, and a regulatory layer consisting of government officials. We introduce PGG to describe the potential behavior of citizens with respect to tax. Specifically, tax compliance is treated as cooperative behavior, while tax evasion is characterized as defection. Unlike the strategies in the game layer, we use ``fair'' and ``corrupt'' to describe the possible behaviors of regulators in the regulatory layer. Citizens in the game layer play the PGG with their neighbors while being supervised by the officials from the regulatory layer. However, not all officials act impartially. Corrupt officials accept bribes from tax evaders and choose not to impose penalties. In contrast, impartial officials will impose punishments on tax evaders according to established rules. Both corrupt and impartial officials can receive supervision fees, which originate from the contribution of taxpayers (who actually produce value). In this way, a corrupt official earns more than a fair partner (as happens in reality). But there is no free lunch, and corrupt officials have other difficulties, such as being more careful not to be revealed, and we can account for this with an additional cost of corruption. Meanwhile, not only do citizens in the game layer update their strategies, but officials in the supervision layer also exhibit dynamic behavioral patterns, tending to imitate more successful neighbors. In this way, we establish the chance of a coevolutionary process where not only strategies but also judgment, meaning the status of supervisors evolve. Methodologically, this research adopts a quantitative and computational modeling approach, employing Monte Carlo (MC) simulations to explore behavioral evolution within taxation systems. Similar quantitative approaches have been used in recent studies of taxation and compliance~\citep{kollruss2025tax, kollruss2025unconstitutionality}. Through MC simulations, we demonstrate that penalties, bribery ratio, regulatory fees, and the costs of corruption all significantly affect tax compliance and fair behaviors in social dilemmas.

The rest of this paper is organized as follows. In Section~\ref{sec2}, we provide a detailed description of our model, including the calculation of payoffs for nodes in both the game layer and the regulatory layer, as well as the strategy update rules. In Section~\ref{sec3}, we outline the simulation methods, present our key findings and analyze their consequences. Finally, we conclude our research and discuss future prospects in Section~\ref{sec4}.

\section{Model}\label{sec2}

In this section, we provide a detailed description of the model, including the network structure, the payoff calculations for both citizens and officials and the strategy update rules.

\subsection{Game Model}

We construct a two-layer network model consisting of a game layer and a regulatory layer, assuming that each layer is a square lattice network with von~Neumann neighborhoods, and nodes interact with their neighbors. In the game layer, we employ PGG to characterize citizens' tax decision-making behaviors. Specifically, each player is involved in multiple PGG organized by the neighbors and the focal player. The positions of citizens and officials in the network are matched, with each official overseeing a PGG centered around their corresponding citizen. This design replicates real-world governance scenarios in which tax officials monitor citizens' compliance behaviors within their jurisdictions. PGG is a typical multi-player social dilemma in which taxpayers contribute some cost to a public pool, while evaders contribute nothing. The total investment in the public pool is then multiplied by a synergy factor $r$ and distributed equally among all citizens, regardless of whether they are taxpayers or evaders.

\subsection{Payoff Calculation}

In this paper, we denote the strategy of citizen (\textit{resp.} regulators) $i$ at time step $t$ as $x_i(t)$ (\textit{resp.} $y_i(t)$). For citizens, $x_i(t)\in\{0,1\}$, where $1$ and $0$ indicates a taxpayer and an evader state, respectively. For regulators, $y_i(t) = 1$ if $i$ is a fair regulator, otherwise $y_i(t)=0$. 

In our model, we assume that the population structure is homogeneous. The payoff for citizen $i$ in PGG organized by its neighbor $j$ (including itself) at time step $t$ can be calculated as:
\begin{equation}\label{1}
\begin{split}
P_{i,j}(t)=
\begin{cases}
    R_j - c x_i(t) - \alpha [1-x_i(t)], 
    & \hfill y_j(t)=1, \\
    R_j - c x_i(t) - \alpha \beta [1-x_i(t)], 
    & \hfill y_j(t)=0.
\end{cases}
\end{split}
\end{equation}
Hereby
\begin{equation}
    R_j = \frac{r \sum\limits_{k \in \Omega_j} c x_k(t)}{\vert \Omega_j\vert}\,,
\end{equation}
 where $\Omega_j$ is the set consisting of the citizen $j$ and its four neighbors, $c$ is the cost contributed by taxpayers to the public pool and $r$ represents the synergy factor. The enhanced contribution $R_j$ is distributed from the public pool to player $j$ and its neighbors. This redistribution mechanism reflects the direct benefits of tax compliance in maintaining public goods. To punish tax evaders, impartial regulators impose a fine of $\alpha\in(1, 2)$ on evaders, whereas corrupt regulators accept bribes from tax evaders to avoid being punished. Since the amount of bribe should be less than the related fine, we set $\beta \in (0, 1)$ as the bribery ratio, making $\alpha \cdot \beta$ the actual bribe that evaders pay to corrupt regulators. Consequently, the total payoff of citizen $i$ at time step $t$ can be calculated as follows:
\begin{equation}\label{2}
    P_i(t) = \sum\limits_{j \in \Omega_i} P_{i,j}(t)\,.
\end{equation}

The payoff for a regulator $i$ at time step $t$ is
\begin{equation}\label{3}
\begin{split}
\Pi_i(t)=
\begin{cases}
    m \sum\limits_{k \in \Omega_i} x_k(t), 
    &
    \\\hfill y_i(t)=1, \\
    m \sum\limits_{k \in \Omega_i} x_k(t) + 
    \beta \sum\limits_{k \in \Omega_i} [1 - x_k(t)] 
    - \gamma, 
    &
    \\  \hfill y_i(t)=0,
\end{cases}
\end{split}
\end{equation}
where $m$ is the regulatory fee and $\gamma$ is the cost of corrupt regulators. For impartial regulators, their income comes from the regulatory fees $m$, which are proportional to the number of taxpayers in PGG. In the real world, law-abiding citizens pay taxes to the government, which uses these funds to maintain social stability, including salaries for government employees. Such system relies fundamentally on high tax compliance rates to sustain its operations. Our model positions corrupt regulators as dual income recipients, they not only receive regulatory fees but also obtain bribes from evaders. However, it does not mean that corrupt regulators can obtain the same regulatory fees as fair regulators, their fees are proportional to the number of tax-compliant participants in the games they oversee. Moreover, corrupt regulators must be more cautious to avoid being discovered for their bribery, which requires them to pay an extra cost of monitoring $\gamma$.

\subsection{Strategy Evolution}

After receiving payoffs, individuals will update their strategies through an imitation process based on pairwise comparison. Importantly, we assume that regulators' strategies are dynamic rather than static, hence they will also update their strategies over time to achieve higher payoffs. This adaptive process drives some originally impartial regulators to abandon their principles and accept bribes from evaders when such behavior proves economically advantageous. Conversely, some corrupt regulators may abandon bribery in subsequent rounds to avoid the extra monitoring costs $\gamma$, resulting in a transition to impartial enforcement. This learning mechanism reflects how tax compliance and fair behaviors spread through social observation of relative benefits. Thus, in both the game layer and regulatory layer, $i$ randomly chooses a neighbor $j$ at each time step and compares their payoffs. The probability that player $i$ adopts the strategy of player $j$ in the next time step is given by the Fermi function:
\begin{equation}\label{4}
    W(S_i \leftarrow S_j)=\frac{1}{1+e^{(P_i - \ P_j) / \kappa}},
\end{equation}
where $P_i$ and $P_j$ represent the strategy and payoff of node $i$ and $j$, respectively. $\kappa$ represents the noise level, reflecting the change of an irrational choice with low probability. It means that individuals are more likely to adopt the strategies of those with higher payoffs than themselves, though stochastic exploration allows for rare irrational strategy updates. As noted, regulators may also change their status and they will update their strategies over time to achieve a higher payoff. The imitation probability is also defined by Eq.~\ref{4}.

\begin{figure*}
    \centering
    \includegraphics[width=0.65\textwidth]{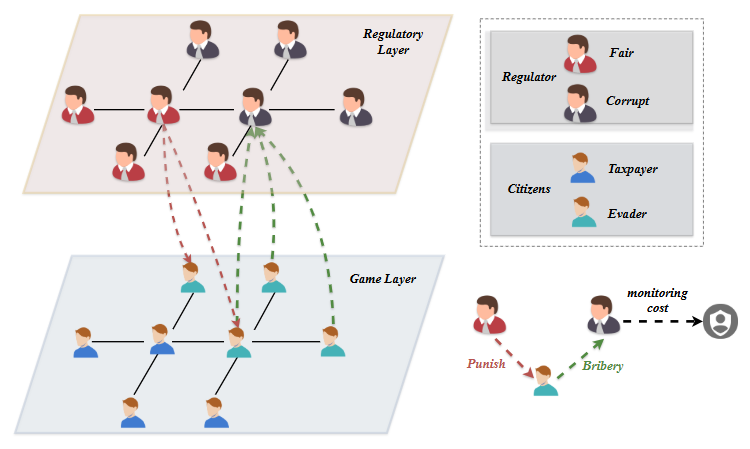}
    \caption{\textbf{Evolutionary game model with regulators.}The lower layer represents the stage of a taxation dilemma, where taxpayers (cooperators) and evaders (defectors) engage in interactions through a spatial public goods game. The upper layer comprises regulators who oversee the evolution of strategies. Fair regulators punish tax evaders, while corrupt regulators accept bribes and help the evaders from being punished. At the same time, to hide their proper behavior, corrupt regulators should pay an additional cost of monitoring.}
    \label{fig1: example}
\end{figure*}
To improve the clarity of our model, we provide a schematic summary of the modeling framework in Fig.~\ref{fig1: example}. As noted, we use a multilayer network approach and divide the population into a game layer and a regulatory layer. The game layer explicitly captures tax compliance dynamics through PGGs, while actors in the regulatory layer supervise these games, hence we establish a connection between the layers. Accordingly, fair regulators impose penalties on evaders in the corresponding group, whereas corrupt regulators accept bribes from evaders and facilitate their evasion of punishment. All regulators receive corresponding regulatory fees that are proportional to the number of taxpayers in their supervised PGGs. However, corrupt regulators must be more cautious to avoid being exposed, as they need to incur additional corruption costs. For strategy updating, we consider that both citizens and regulators update their strategies via an imitation process based on pairwise comparison of payoff values.

\section{Simulation results and analyses}\label{sec3}

In this section, we introduce simulated methods in detail, and analyze the coevolution of the game layer and the regulatory layer, demonstrating how the citizens' tax compliance rate and the proportion of fair regulators are influenced by the key parameters including the penalty $\alpha$, the bribery ratio $\beta$, the regulatory fees $m$ and the cost of monitoring $\gamma$. Finally, we demonstrate the strong correlation of pattern formations between citizens and regulators.

\subsection{Methods}

Our approach is fully quantitative, relying on numerical simulations. In the initial stage of each simulation, citizens and regulators are embedded into two $50\times50$ square-lattice networks with periodic boundaries. Each individual has an equal probability of being a taxpayer (fair regulator) or a tax evader (corrupt regulator). During a MC simulation step, on average, all individuals are randomly selected to compare a neighboring payoff, and decide whether to adopt that neighbor's strategy with a probability $W$ given in Eq.~\ref{4}. Based on our simulations, evolution reaches the stationary state after 2000 steps. Therefore, we average the last 500 steps of 3000 total steps to calculate the tax compliance rate and proportion of fair regulators. Each result is averaged over 10 simulations to ensure accuracy. For simplicity, we set $c = 1$ and $ \kappa = 0.1$.

\subsection{Effect of penalty $\alpha$ on tax compliance and fairness}

\begin{figure*}[t]
    \centering
    \begin{subfigure}{0.35\textwidth}
        \centering
        \includegraphics[width=\textwidth]{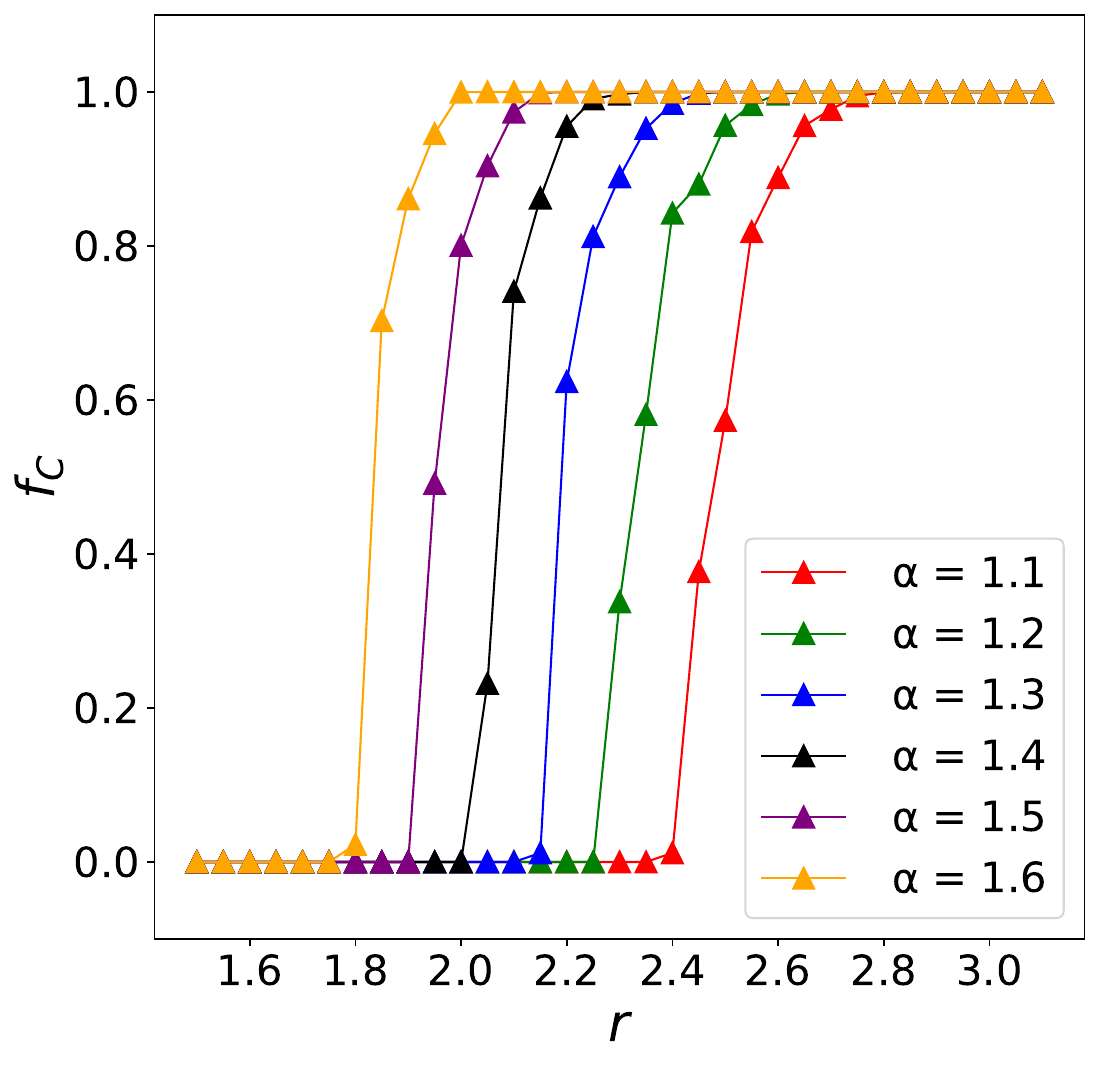}
        \caption{$f_C$}
        \label{fig2:a}
    \end{subfigure}
    %\hfill
    \hspace{-1mm}
    \begin{subfigure}{0.35\textwidth}
        \centering
        \includegraphics[width=\textwidth]{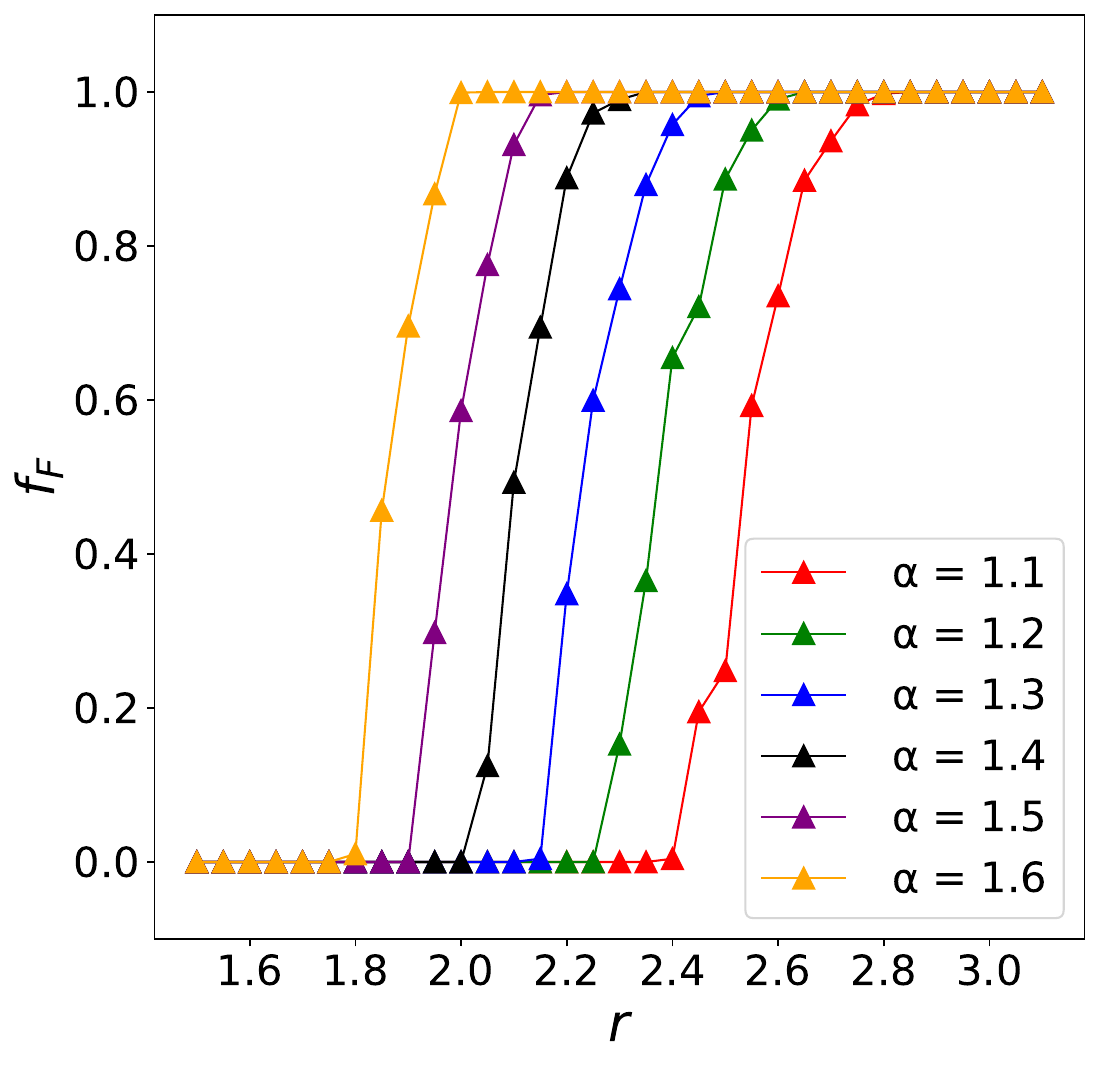}
        \caption{$f_F$}
        \label{fig2:b}
    \end{subfigure}
    \caption{\textbf{Tax compliance rate and fair regulators density as a function of synergy factor at different fine values.} Panel~(a) illustrates the proportion of taxpayers in the game layer as the synergy factor $r$ increases, with the values of fine $\alpha$ for evaders marked in the legend. Panel~(b) displays the proportion of fair regulators in the regulatory layer. We set the bribery ratio $\beta=0.3$, the supervision fee $m=0.1$ and the cost of monitoring $\gamma=0.1$. Each data point is averaged from the last 500 steps of 3000 total steps and averaged over 10 independent simulations.} 
    \label{fig2}
\end{figure*}

To characterize system behavior, we introduce two new metrics, $f_C$ and $f_F$, which quantify the frequencies of compliant taxpayers in the game layer and fair regulators in the regulatory layer, respectively, enabling simultaneous analysis of both behavioral and institutional dimensions. 

We first focus on the impact of penalty $\alpha$ on evolutionary behavior and the curves of $f_C$ and $f_F$ against the synergy coefficient $r$ for different $\alpha$ values in Fig.~\ref{fig2}. As shown in this figure, the system exhibits a typical phase transition: at small $r$ values, both the game layer and the regulatory layer remain in states of pure tax evasion and pure corruption, respectively. As $r$ increases, the system undergoes a rapid transition, ultimately reaching states of pure tax compliance and pure fair regulation. Generally, $\alpha=1.6$ provides the optimal parameter space for pure compliance and pure fairness states, while $\alpha=1.1$ offers the most favorable conditions for pure evasion and pure corruption. For $\alpha=1.1$, the threshold for pure evasion and pure corruption is about $r=2.37(2)$, representing the highest values observed. At $\alpha=1.2$ and $\alpha=1.3$, the thresholds for the emergence of taxpayers are $r=2.25$ and $r=2.1$, while those for fair regulators are $r=2.25$ and $r=2.15$, respectively. When $\alpha=1.4$ and $\alpha=1.5$, the existence thresholds for taxpayers in the game layer are $r=2.0$ and $r=1.9$, with identical thresholds observed for impartial regulators in the regulatory layer. These results clearly demonstrate that the penalty parameter $\alpha$ plays a crucial regulatory role in the evolutionary game: higher punishment levels can lower the threshold for the emergence of tax compliance. When $\alpha=1.6$, tax-compliant behavior begins to emerge in the game layer at a synergy factor of just $r=1.75$, while fair regulatory behavior appears simultaneously in the regulatory layer. As $r$ continues to increase to 2.0, the system rapidly converges to an ideal equilibrium state of global pure compliance and pure fairness. This is because when $\alpha$ is sufficiently large, evaders are required to pay higher fines, making it easier for taxpayers to gain an advantage in PGG. Crucially, the fair enforcement in the regulatory layer and tax-compliant behavior in the game layer mutually reinforce each other, creating a positive feedback cycle. The high fine $\alpha$ serves dual functions: it directly suppresses evasion while indirectly promoting regulatory fairness. 

These findings provide important insights for designing effective tax compliance policies: in real social systems, increasing punishment levels appropriately can more effectively reduce tax evasion among citizens.

\begin{figure*}[h]
    \centering
    \begin{subfigure}{0.35\textwidth}
        \centering
        \includegraphics[width=\textwidth]{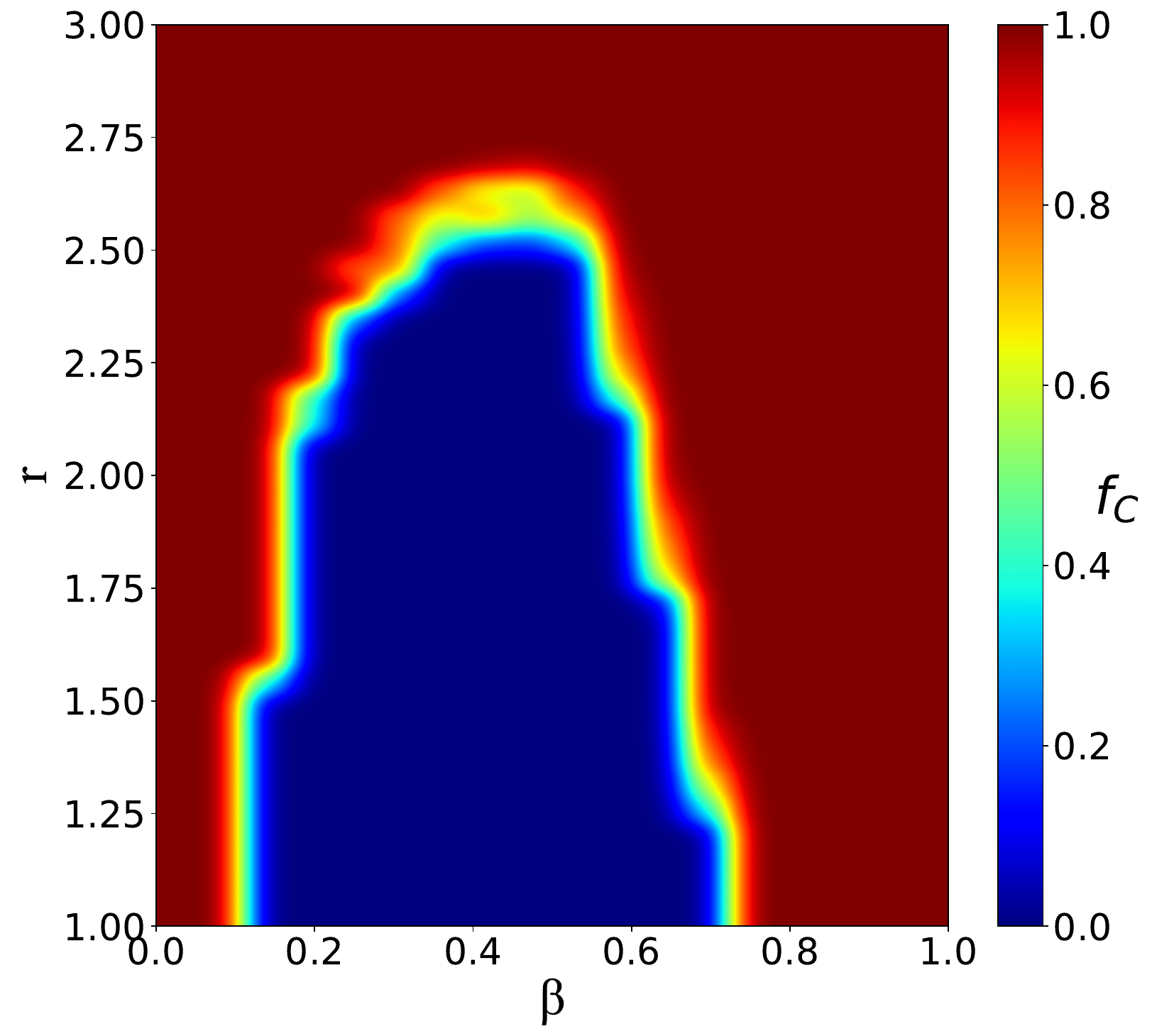}
        \caption{$f_C$}
        \label{fig3:a}
    \end{subfigure}
    \hspace{-1mm}
    \begin{subfigure}{0.35\textwidth}
        \centering
        \includegraphics[width=\textwidth]{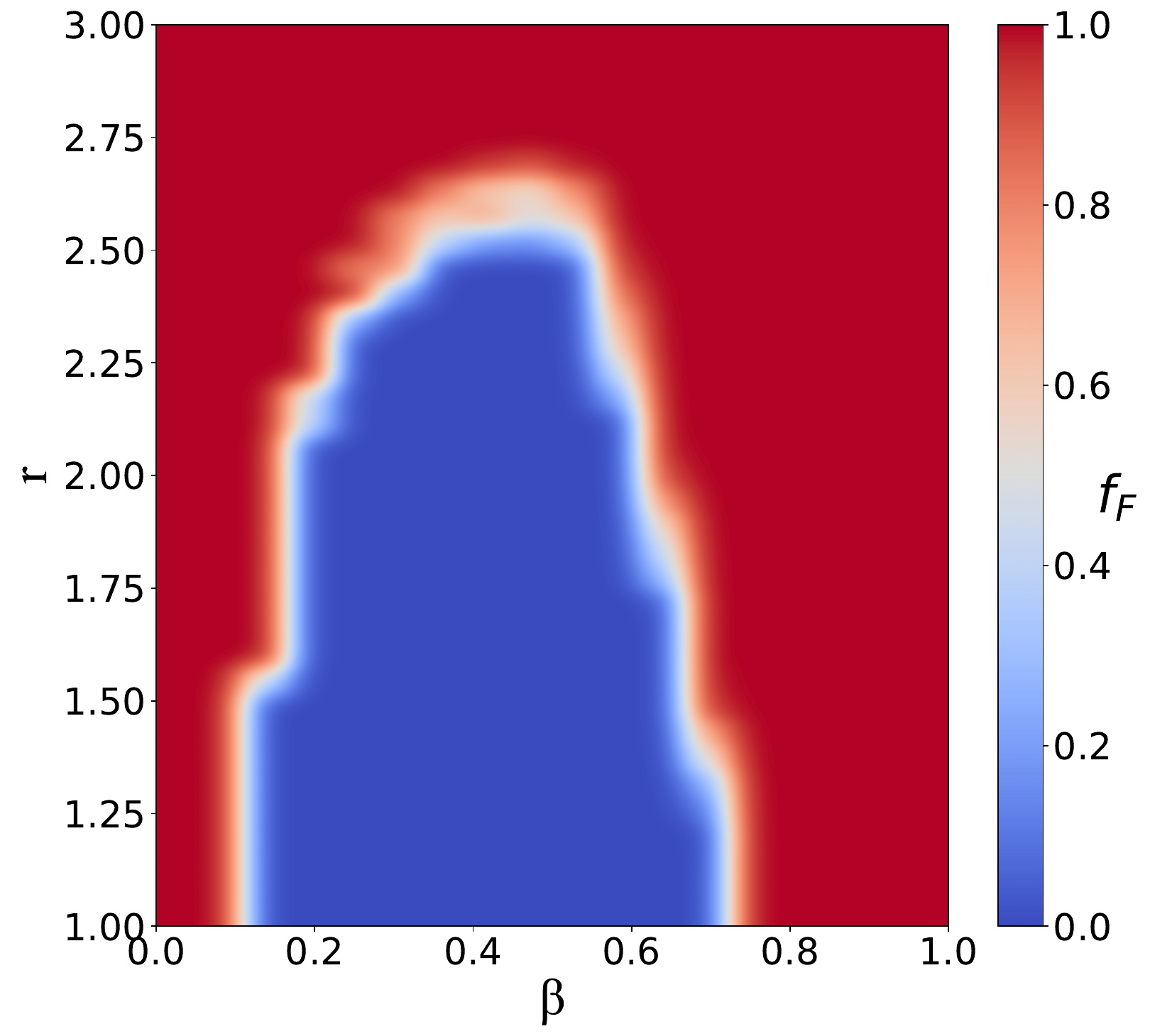}
        \caption{$f_F$}
        \label{fig3:b}
    \end{subfigure}
    \caption{\textbf{The tax compliance rate and the proportion of fair regulators in dependence of synergy factor and bribery ratios.} The left panels show the color-coded portion of taxpayers in the game layer on the $\beta-r$ parameter plane. The right panels show the color-coded density of fair regulators in the regulatory layer at the same $\beta-r$ parameter pairs. The remaining parameters, the penalty $\alpha=1.2$, the supervision fee $m=0.1$ and the cost of monitoring $\gamma=0.1$ are fixed. All results are averaged from the last 500 steps of 3000 total steps and averaged over 10 independent simulations.} 
    \label{fig3}
\end{figure*}

\subsection{Effect of bribery ratio $\beta$ on the $f_C$ and $f_F$}

Next, we focus on the effect of the bribery ratio $\beta$ on individual behavior. Figure~\ref{fig3} presents the $\beta$-$r$ parameter space heatmaps for both the tax compliance rate $f_C$ and the portion of fair regulators $f_F$. The results demonstrate that $\beta$ exerts a nonlinear influence on both the tax compliance rate $f_C$ and the proportion of fair regulators $f_F$. When $\beta<0.06$, taxpayers and fair regulators dominate the entire network. This occurs because the bribes given by evaders to corrupt regulators become too low to be sustainable. Although the number of evaders increases, the total bribes for corrupt regulators remain insufficient to compensate for their risk exposure. Furthermore, as the supervised PGGs contain too few taxpayers, the collected supervision fees decrease significantly. Consequently, corrupt regulators ultimately earn lower payoffs than fair partners, leading to the complete dominance of fair regulators in the regulatory layer. For evaders in the game layer, they can only pay high fines due to their tax-evasion behavior, which ultimately allows taxpayers to occupy the entire game layer. As $\beta$ increases, the thresholds $r$ required to reach the pure compliance and pure fairness states also rise. When $\beta$ reaches 0.45, the threshold $r$ reaches its peak. This phenomenon occurs because as bribery increases, corrupt regulators receive sufficient bribes to offset their risk exposure, leading to their eventual dominance in the regulatory layer. At this point, evaders in the game layer only need to pay a small bribe to escape high penalties, resulting in evaders ultimately dominating the entire network. 

However, when $\beta>0.45$, the bribery ratio shows a negative correlation with the synergy coefficient $r$. Specifically, when $\beta$ exceeds 0.75, taxpayers and fair regulators continue to dominate the network even at low $r$ values. This is because the continuously growing bribery ratio forces evaders to pay increasingly higher bribes, ultimately making tax compliance the more economically rational choice. Corrupt regulators lose their bribery income and consequently earn less than fair regulators. Consequently, they become fair regulators in the next round. 

In real-world governance, government agencies should focus on the ratio of fines to the amount of bribery received by officials, establishing a dynamic tiered supervision system. This system should adapt anti-corruption strategies in real-time based on $\beta$ values to foster clean governance, thereby enabling more effective regulation of citizens' tax compliance behavior.

\begin{figure*}[t]
    \centering
    \begin{subfigure}[b]{0.35\textwidth}
        \centering
        \includegraphics[width=\textwidth]{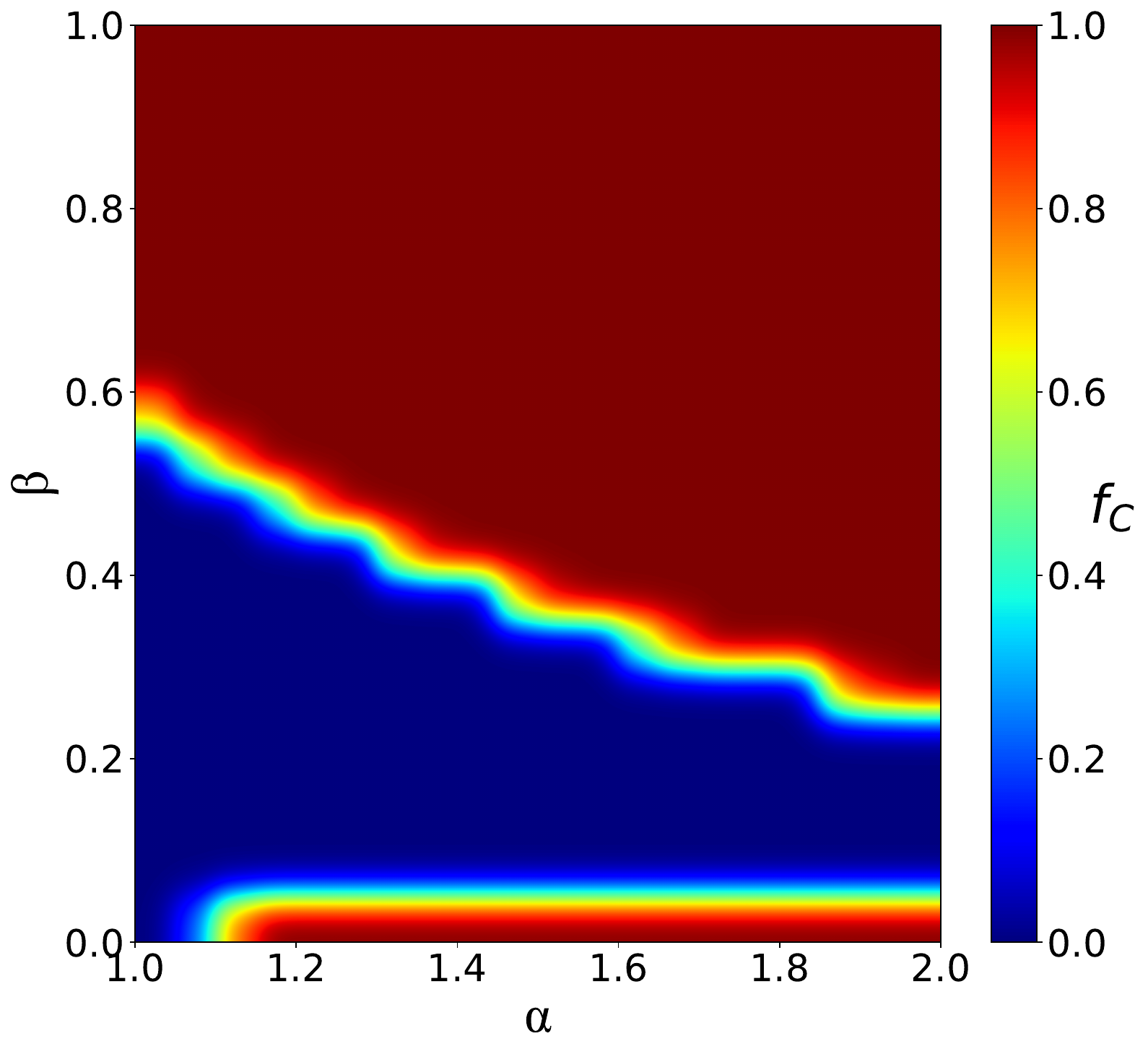}
        \caption{$f_C$ at $m=0.1$, $\gamma=0.1$}
        \label{fig4:a}
    \end{subfigure}
    \hspace{-3mm}
    \begin{subfigure}[b]{0.35\textwidth}
        \centering
        \includegraphics[width=\textwidth]{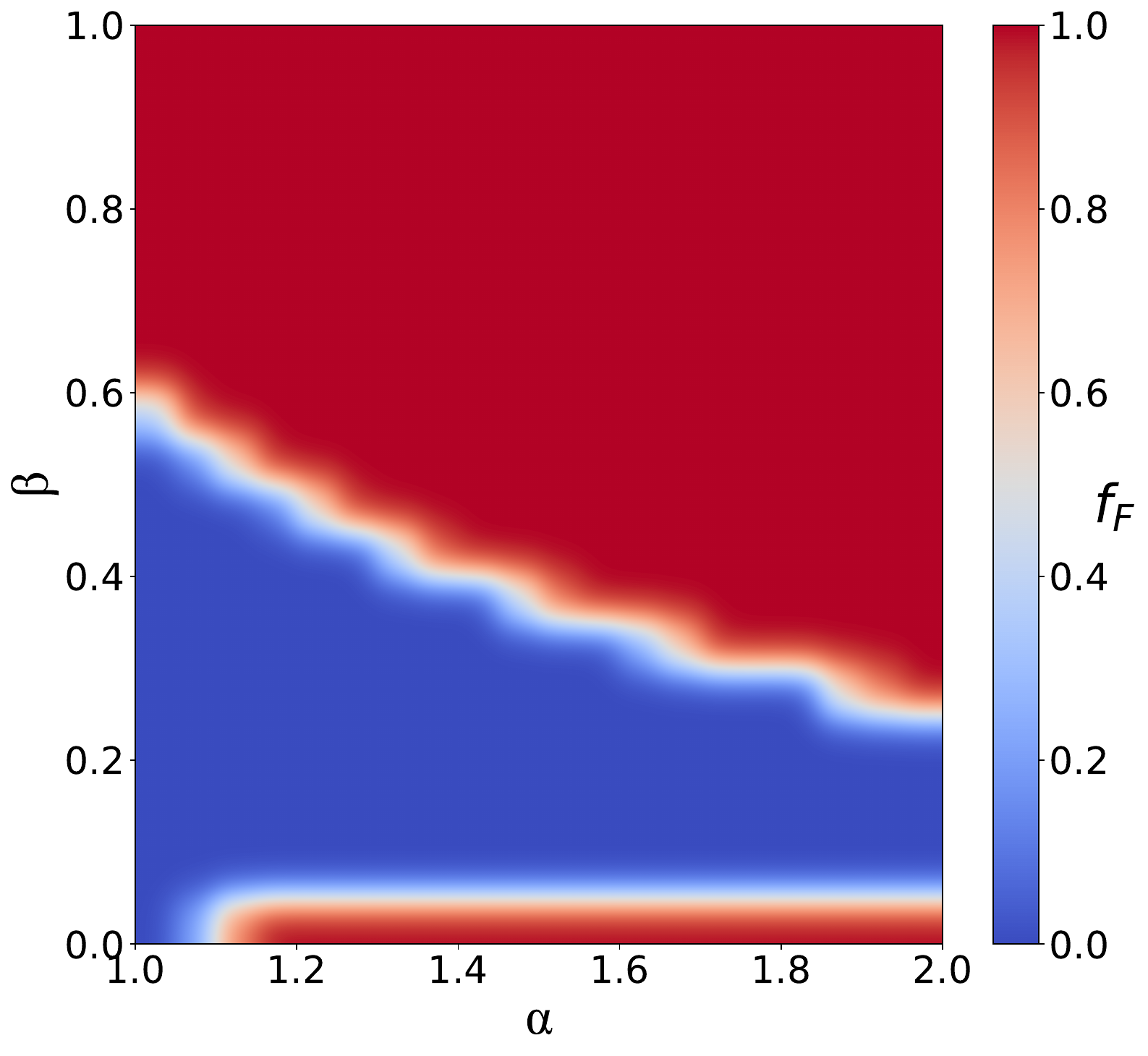}
        \caption{$f_F$ at $m=0.1$, $\gamma=0.1$}
        \label{fig4:b}
    \end{subfigure}

    \begin{subfigure}[b]{0.35\textwidth}
        \centering
        \includegraphics[width=\textwidth]{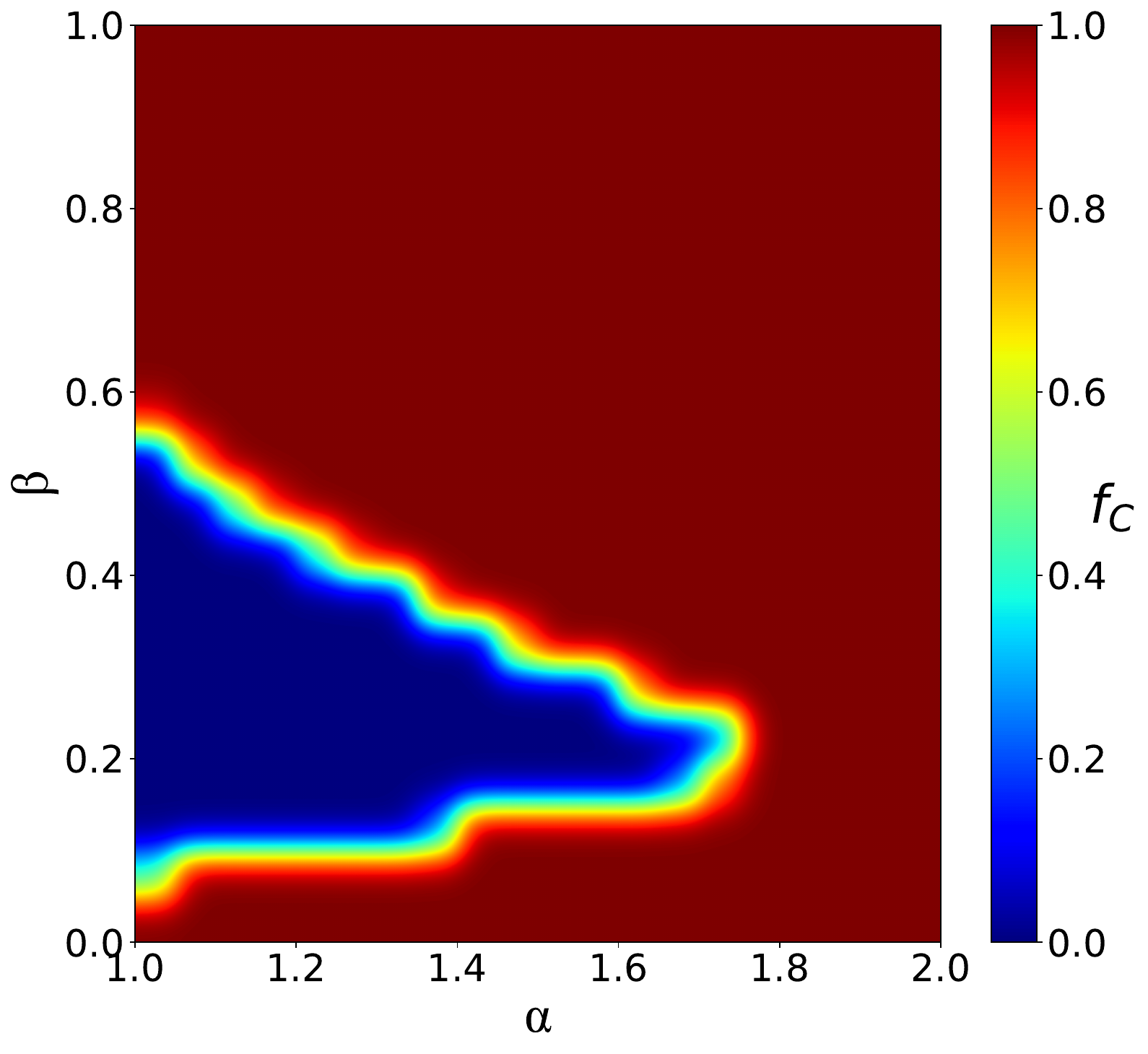}
        \caption{$f_C$ at $m=0.2$, $\gamma=0.2$}
        \label{fig4:c}
    \end{subfigure}
    \hspace{-3mm}
    \begin{subfigure}[b]{0.35\textwidth}
        \centering
        \includegraphics[width=\textwidth]{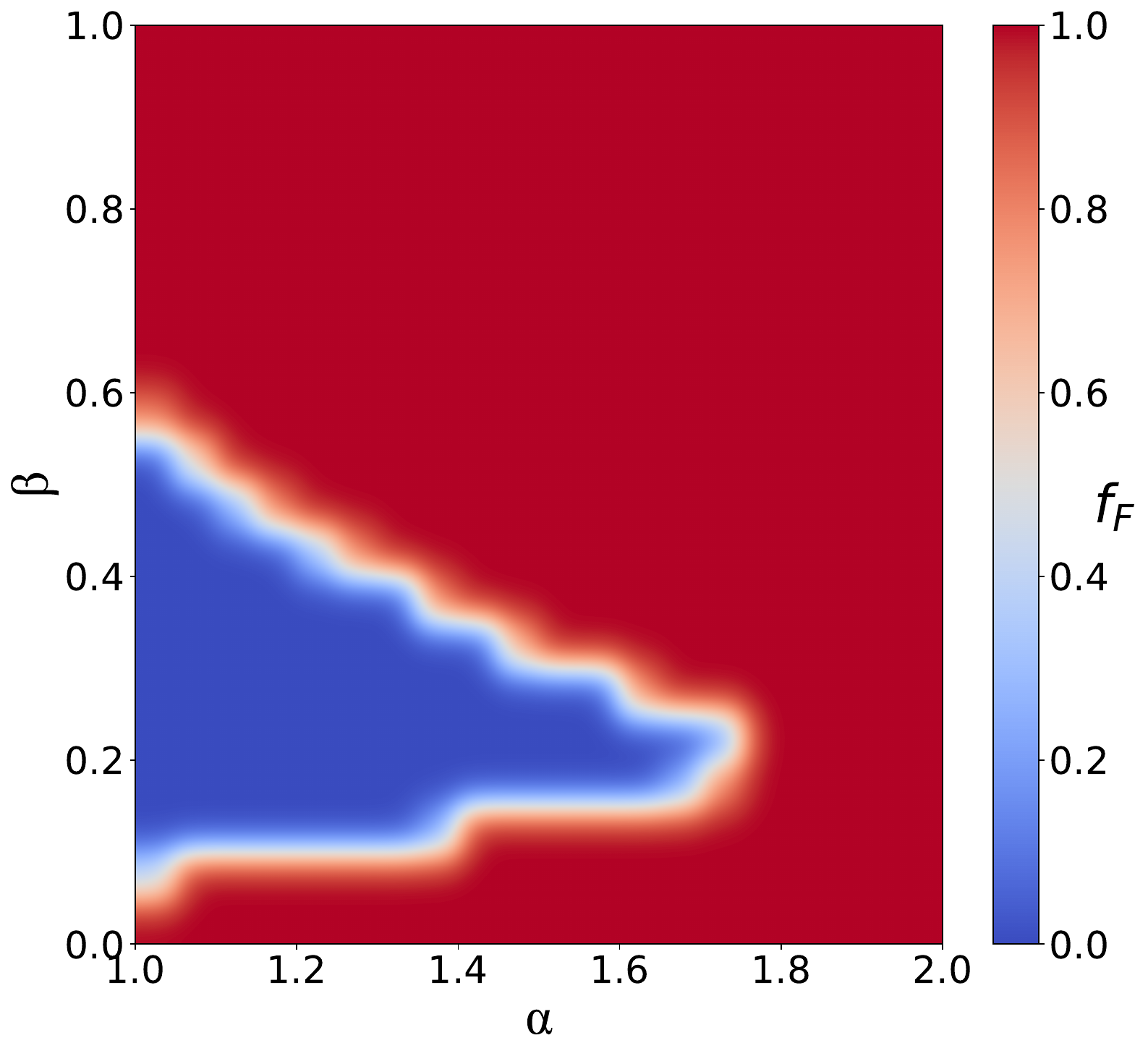}
        \caption{$f_F$ at $m=0.2$, $\gamma=0.2$}
        \label{fig4:d}
    \end{subfigure}
    \vspace{-1mm}
    
    \caption{\textbf{The influence of parameter pairs ($\alpha$, $\beta$) on individual behaviors.} The left panels show the color-coded portion of taxpayers in the game layer on the $\alpha-\beta$ parameter plane. Right panels show the color-coded density of fair regulators in the regulatory layer at the same parameter pairs. The top row shows the case obtained at supervision fee $m=0.1$ and corruption cost $\gamma=0.1$, while the bottom row indicates when $m=0.2$ and $\gamma=0.2$. The synergy factor is fixed at $r=1.5$ for all cases. The stationary values are calculated from the last 500 steps of 3000 total steps and averaged over 10 independent simulations.}
    \label{fig4}
\end{figure*}

\begin{figure*}[t]
    \centering
    \begin{subfigure}{0.35\textwidth}
        \centering
        \includegraphics[width=\textwidth]{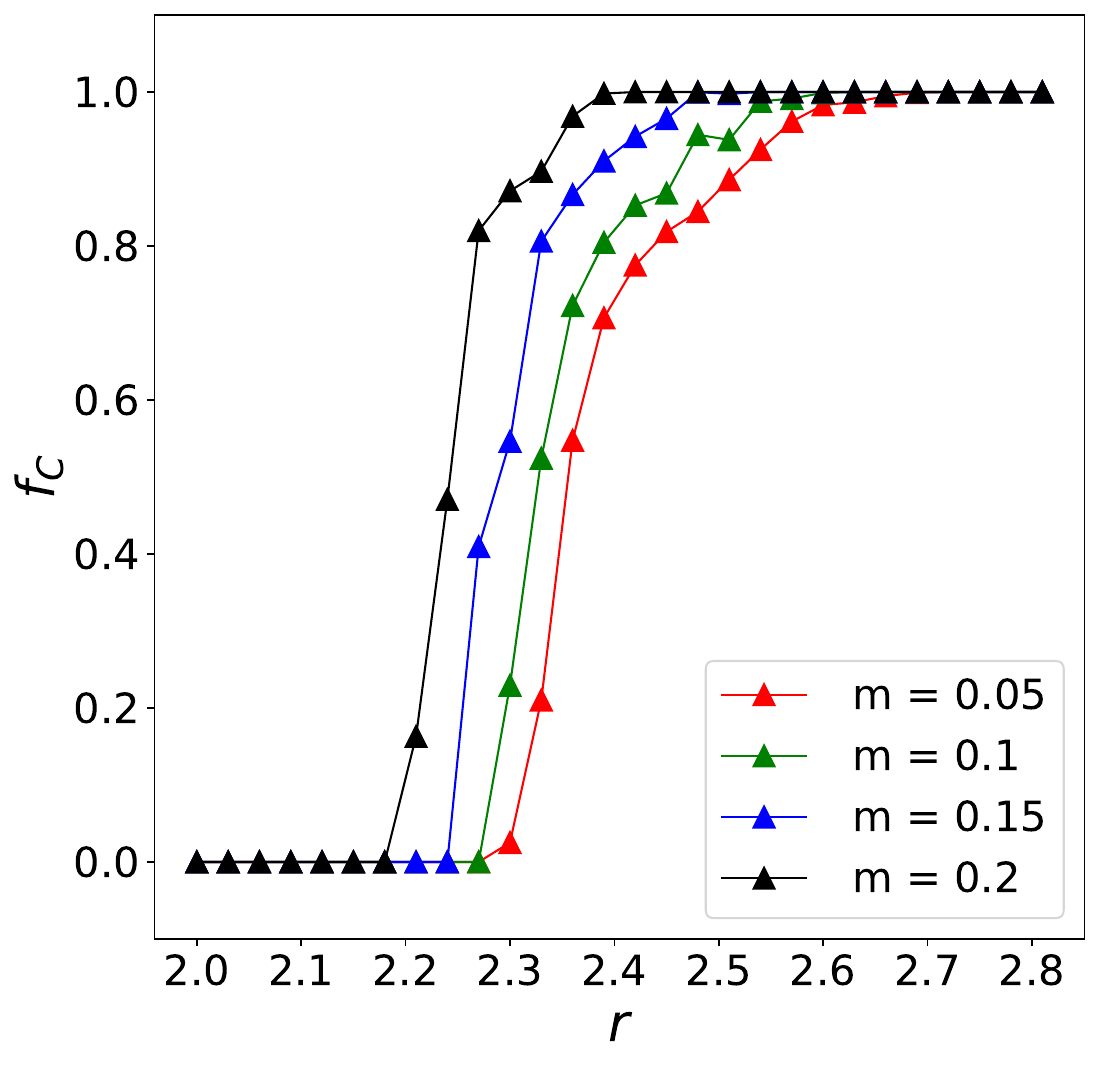}
        \caption{$f_C$}
        \label{fig5:a}
    \end{subfigure}
    \hspace{-1mm}
    \begin{subfigure}{0.35\textwidth}
        \centering
        \includegraphics[width=\textwidth]{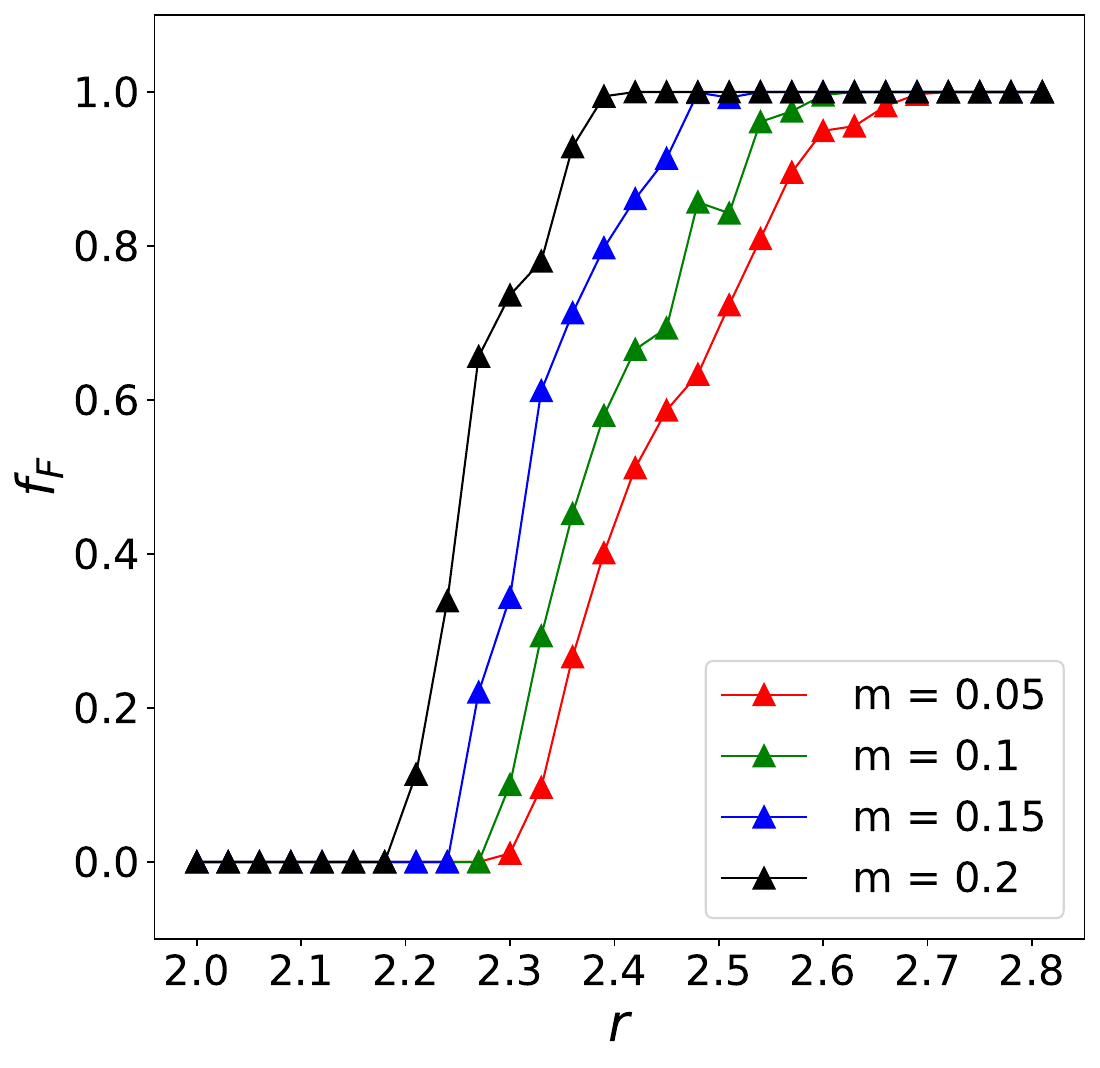}
        \caption{$f_F$}
        \label{fig5:b}
    \end{subfigure}
    \caption{\textbf{Degree of tax compliance and fairness level among regulators in dependence of synergy factor at different regulatory fees. }Panel~(a) shows the fraction of taxpayers in the game layer as we increase the synergy factor $r$, while panel~(b) depicts the fraction of fair regulators in the regulatory layer. The values of regulatory fees that all regulators can obtain are marked in the legend. We set the penalty $\alpha=1.2$, the bribery ratio $\beta=0.3$, and the cost of corruption $\gamma=0.1$. Each result is obtained by averaging the last 500 steps out of 3000 total steps, and the average is taken over 10 simulations.} 
    \label{fig5}
\end{figure*}

\subsection{The landscapes of tax compliance and regulator fairness on punishment-bribe parameter space}

To explore the combined effects of the penalty $\alpha$ and the bribery ratio $\beta$ on the evolution more generally, we summarize the key metrics on the parameter space depicted in Fig.~\ref{3}. Panels~(a) and (b) present the values of $f_C$ and $f_F$ at parameter conditions $m=0.1$ and $\gamma=0.1$, while panels~(c) and (d) display the corresponding results for $m=0.2$ and $\gamma=0.2$. Generally, tax compliance rate and fair regulators density exhibit similar trends on the heatmap. 

As a general observation, both $f_C$ and $f_F$ grow when we increase the punishment ratio or the bribery cost when $\beta>0.22$. When $\beta<0.22$, higher penalties or lower bribery ratios are more conducive to the survival of taxpayers and fair regulators. For a specified penalty $\alpha$, when the bribery ratio $\beta$ is sufficiently small, both $f_C$ and $f_F$ can easily reach 1 if we slightly increase the value of $\alpha$. As $\beta$ increases, $f_C$ and $f_F$ initially decrease and then increase again. For each specified $\beta$, as $\alpha$ increases, both $f_C$ and $f_F$ rise. All of these are consistent with our previous analysis. Moreover, when $\beta<0.22$, the top row and the bottom row exhibit different results, which are mainly caused by the regulatory fee $m$ and the cost of corruption $\gamma$. Both $m$ and $\gamma$ in Fig.~\ref{fig4:c} and Fig.~\ref{fig4:d} are greater than those in Fig.~\ref{fig4:a} and Fig.~\ref{fig4:b}. The higher regulatory fee allows fair regulators to better resist temptation, while the greater cost of corruption means that corrupt regulators need to collect more bribes to offset their risk exposure. Therefore, the blue area of the top row is larger than that in the bottom row. The effects of regulatory fee $m$ and monitoring cost $\gamma$ on the degree of tax compliance and the intensity of free regulation will be discussed in detail in the following section.

In real-world tax governance practice, government need to establish a ``dual disciplinary'' mechanism: on one hand, they must continuously strengthen rigid constraints on tax evaders by dynamically adjusting penalty amounts and implementing joint disciplinary measures to increase the cost of non-compliance; on the other hand, they must simultaneously enhance institutional controls against bribery among tax officials. This dual governance model can effectively maintain the synergistic growth of $f_C$ and $f_F$, achieving a virtuous cycle in the tax ecosystem.

\begin{figure*}[t]
    \centering
    \begin{subfigure}{0.35\textwidth}
        \centering
        \includegraphics[width=\textwidth]{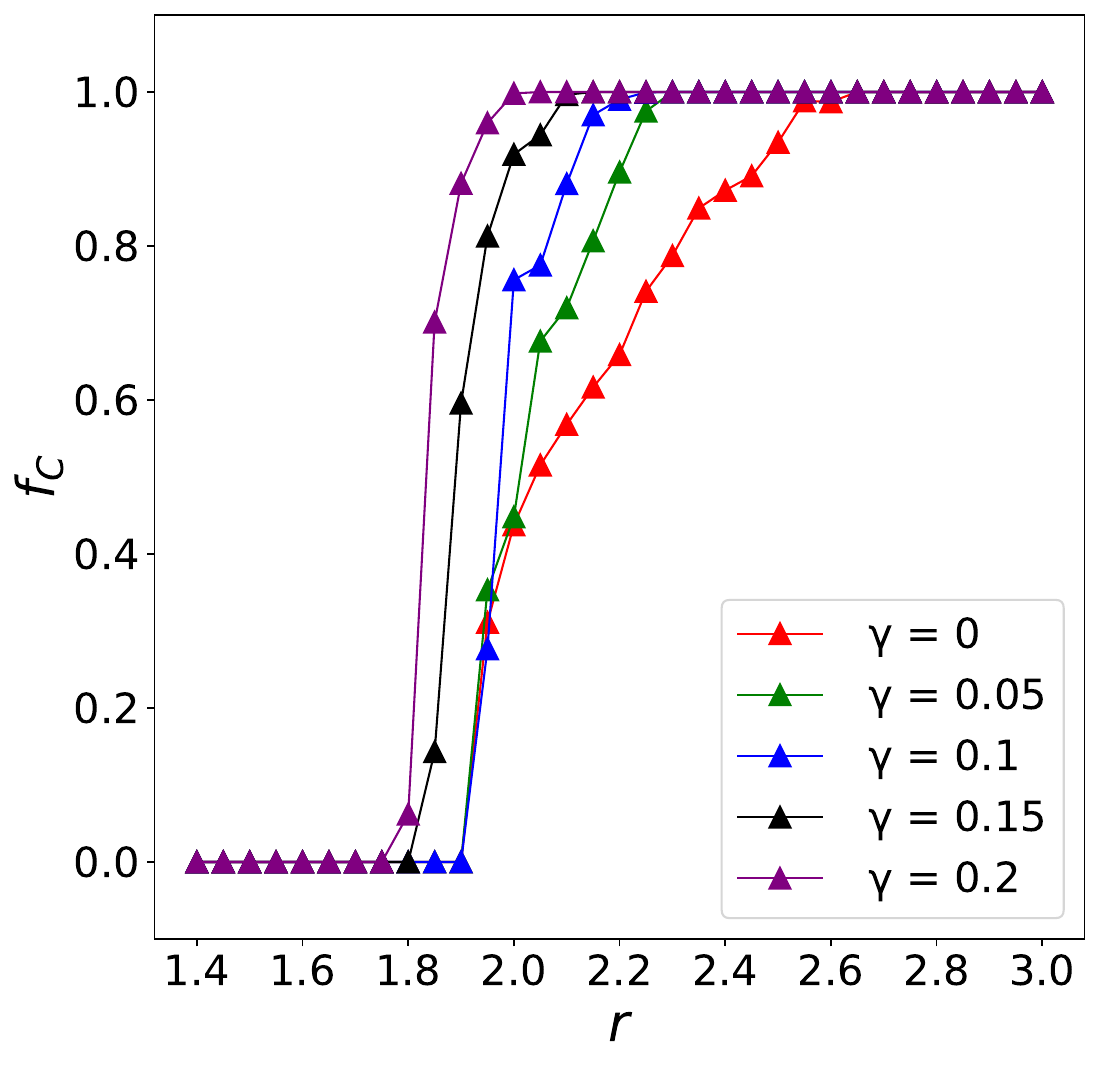}
        \caption{$f_C$}
        \label{fig6:a}
    \end{subfigure}
    \hspace{-1mm}
    \begin{subfigure}{0.35\textwidth}
        \centering
        \includegraphics[width=\textwidth]{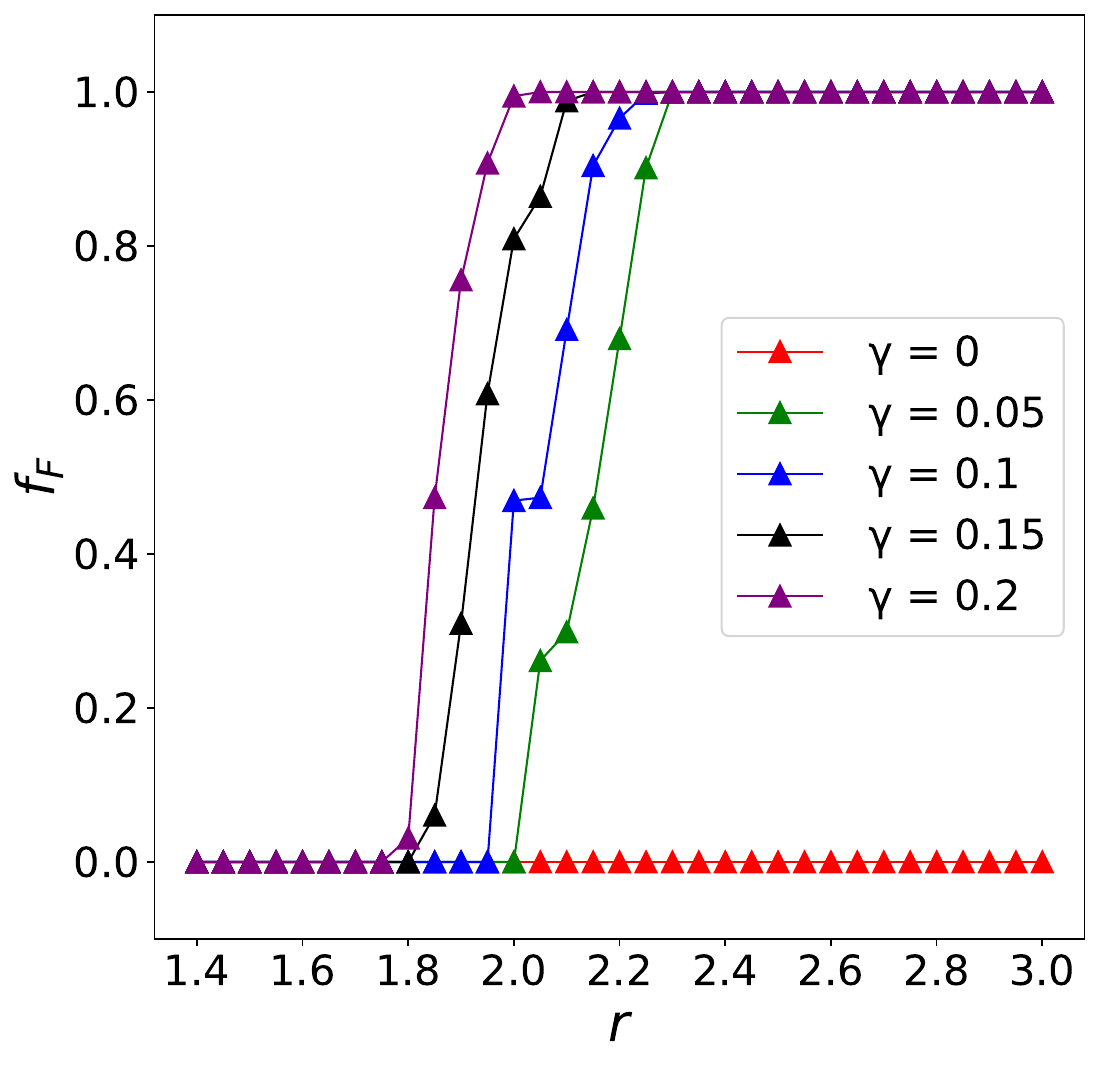}
        \caption{$f_F$}
        \label{fig6:b}
    \end{subfigure}
    \caption{\textbf{Tax compliance and regulation fairness against the synergy factor at different monitoring costs.} Figures~(a) and (b) respectively illustrate the results of the tax compliance rate in the game layer and the proportion of fair regulators in the regulatory layer as the synergy factor $r$ increases. The values of the monitoring cost are marked in the legend. We set the penalty $\alpha=1.2$, the bribery ratio $\beta=0.4$, and the regulatory fee $m=0.1$. The stationary values are calculated from the last 500 steps of 3000 total steps and averaged over 10 independent simulations.} 
    \label{fig6}
\end{figure*}

\begin{figure*}[t]
    \centering
    \begin{subfigure}{0.35\textwidth}
        \centering
        \includegraphics[width=\textwidth]{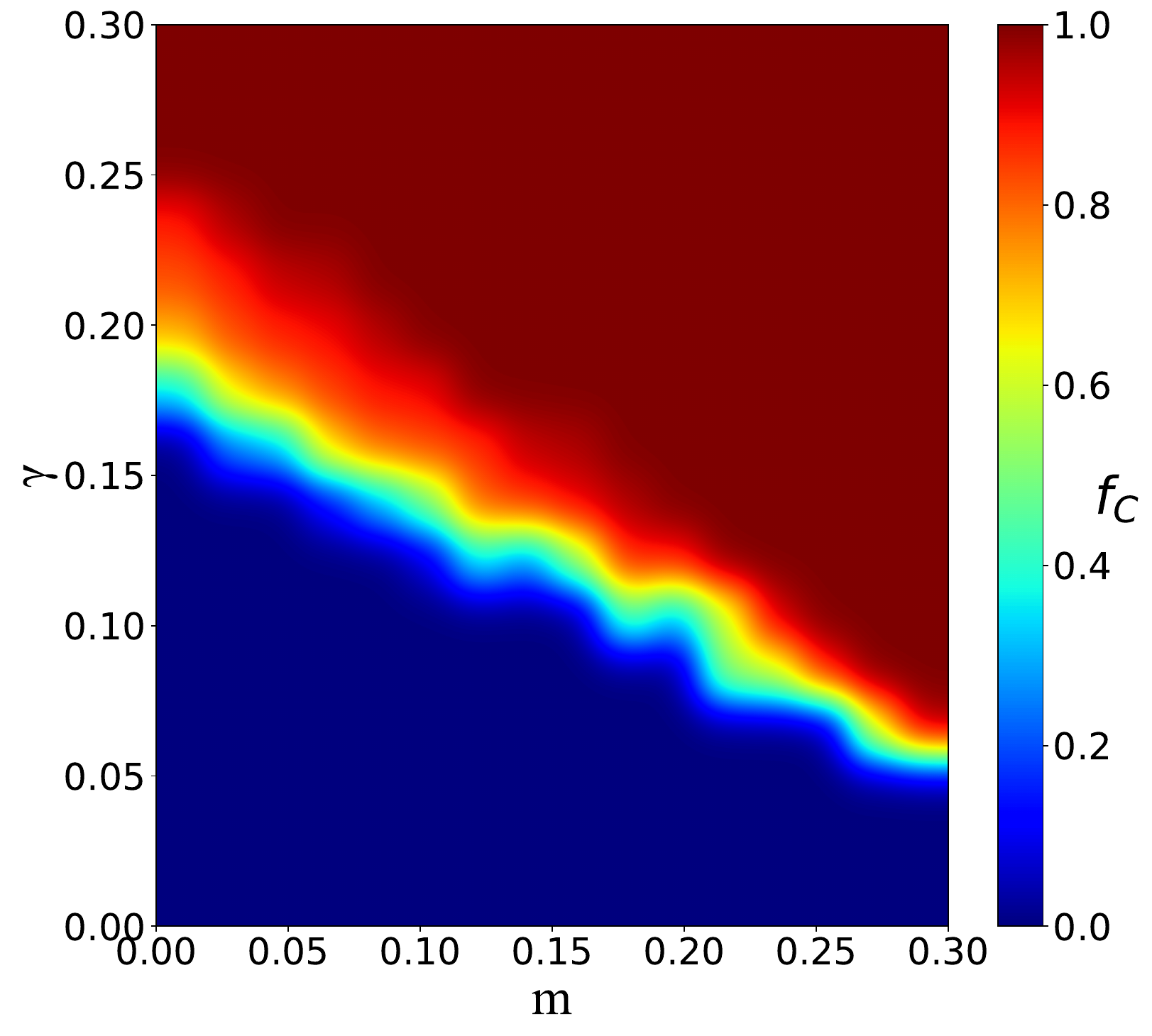}
        \caption{$f_C$ at $\alpha=1.4$, $\beta=0.3$}
        \label{fig7:a}
    \end{subfigure}
    \hspace{-3mm}
    \begin{subfigure}{0.35\textwidth}
        \centering
        \includegraphics[width=\textwidth]{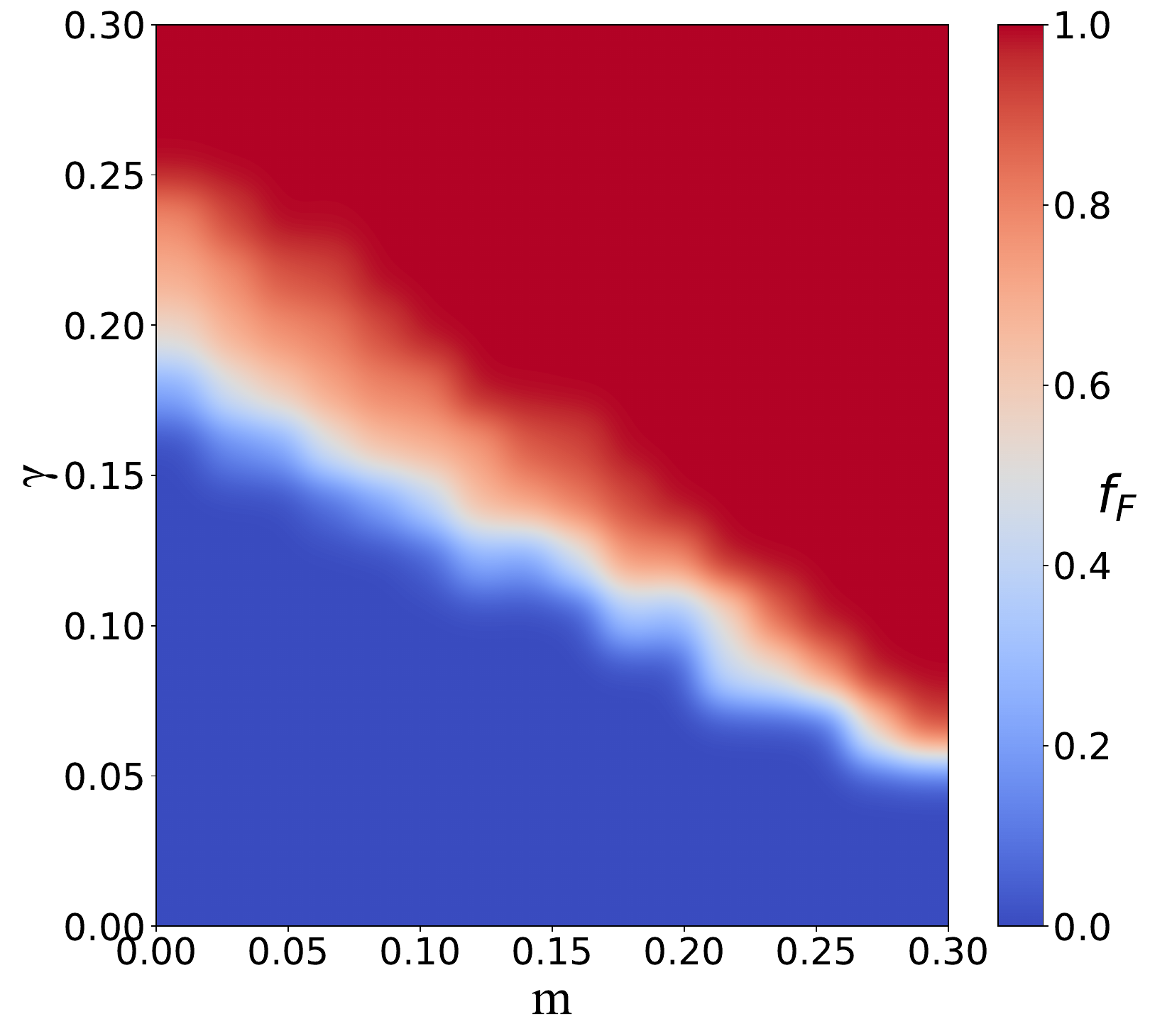}
        \caption{$f_F$ at $\alpha=1.4$, $\beta=0.3$}
        \label{fig7:b}
    \end{subfigure}

    \begin{subfigure}{0.35\textwidth}
        \centering
        \includegraphics[width=\textwidth]{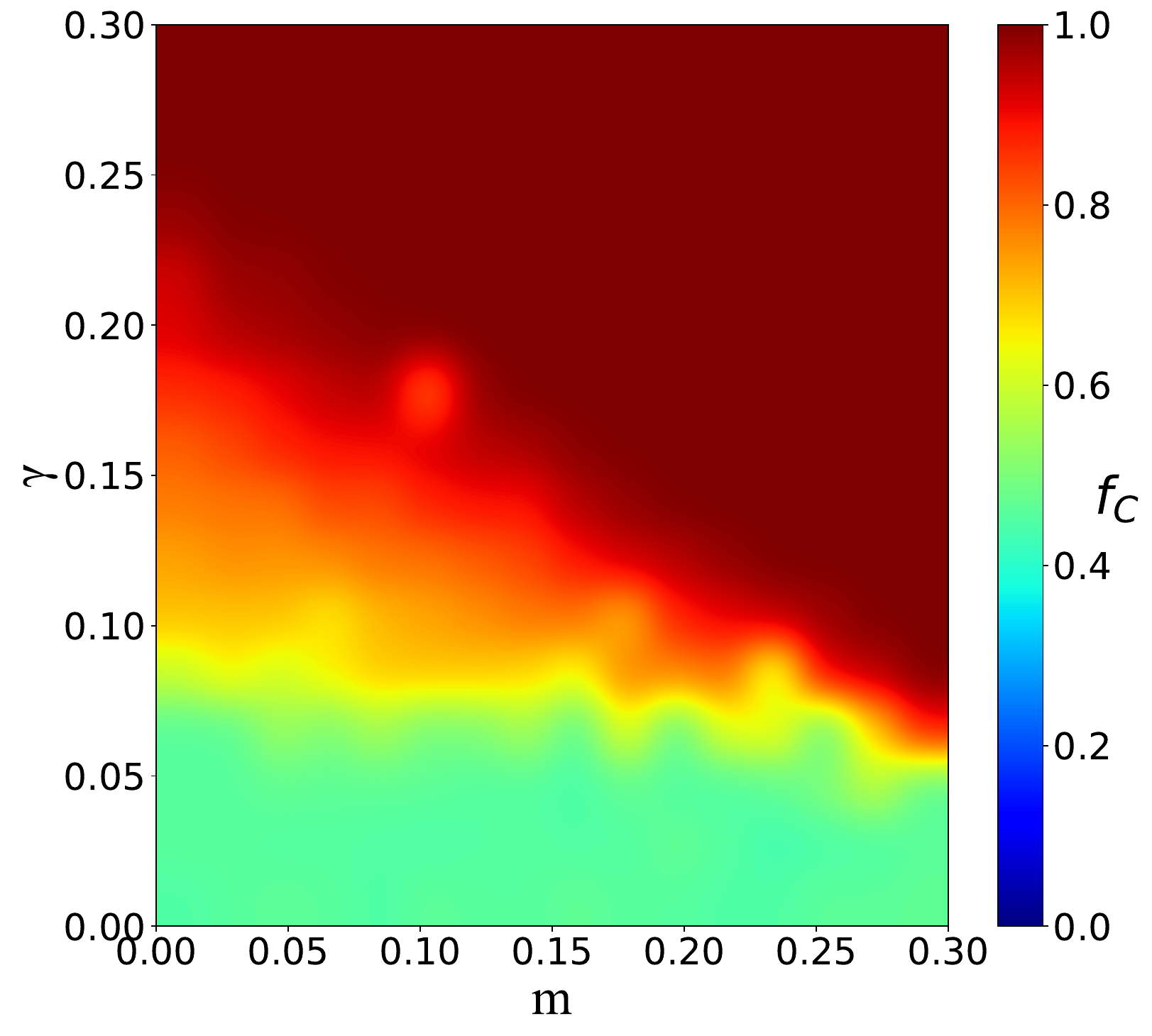}
        \caption{$f_C$ at $\alpha=1.2$, $\beta=0.4$}
        \label{fig7:c}
    \end{subfigure}
    \hspace{-3mm}
    \begin{subfigure}{0.35\textwidth}
        \centering
        \includegraphics[width=\textwidth]{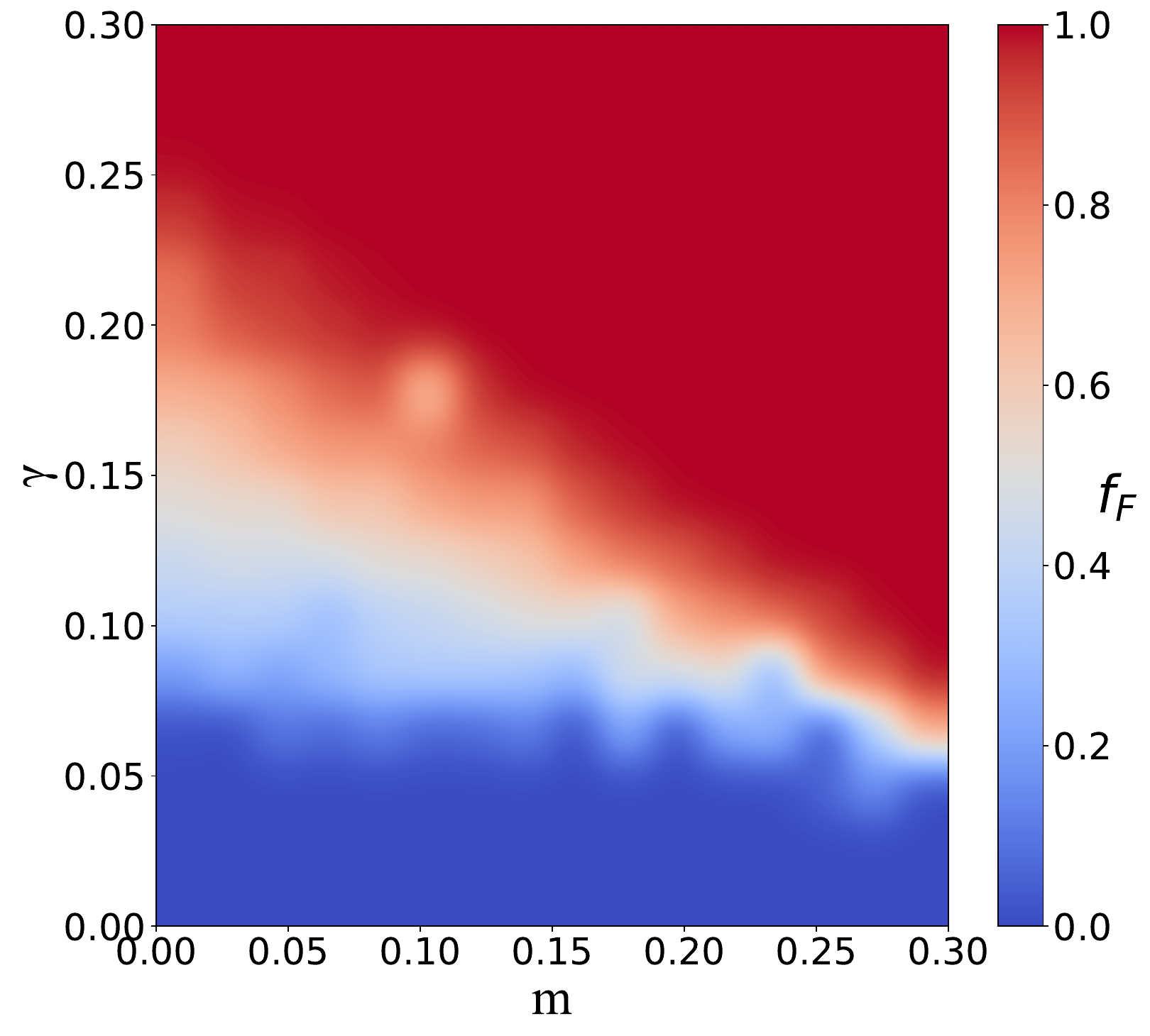}
        \caption{$f_F$ at $\alpha=1.2$, $\beta=0.4$}
        \label{fig7:d}
    \end{subfigure}
    \vspace{-1mm}
    
    \caption{\textbf{The influence of parameter pairs ($m$, $\gamma$) on individual behaviors.} Panels~(a) and (c) show the color-coded portion of taxpayers in the game layer on $m-\gamma$ parameter plane. Panels~(b) and (d) show the color-coded density of fair regulators in the regulatory layer at the same parameter pairs. The top row shows the case obtained at penalty $\alpha=1.4$ and bribery ratio $\beta=0.3$, while the bottom row indicates when $\alpha=1.2$ and $\beta=0.4$. The synergy factor is fixed at $r=2.0$ for all cases. The $f_C$ and $f_F$ for each data point are averaged from the last 500 steps of 3000 total steps and averaged over 10 independent simulations.}
    \label{fig7}
\end{figure*}

\subsection{The effect of regulatory fee $m$ on $f_C$ and $f_F$}

To focus on the impact of regulatory cost $m$ on individuals’ behaviors, we plot the curves of tax compliance rate $f_C$ and fairness ratio $f_F$ against the synergy factor $r$ at different values of $m$. We find that both $f_C$ and $f_F$ exhibit roughly the same trend when other parameters are fixed in Fig.~\ref{fig5}. 

From Fig.~\ref{fig5}, we observe that both $f_C$ and $f_F$ increase with the synergy factor $r$. Additionally, higher values of the regulatory cost $m$ correspond to lower existence thresholds for evaders and corrupt regulators. For $m=0.2$, the threshold values for both taxpayers and fair regulators are $r=2.18$. In contrast, for $m=0.05$, taxpayers and fair regulators do not appear until $r$ reaches 2.3. The payoff for fair regulators depends on both the supervision fee $m$ and the number of taxpayers in their supervised PGGs. In addition to the factors mentioned above, the payoff for corrupt regulators is also related to the amounts of bribes from evaders and the costs associated with their corruption. Generally, the PGG supervised by fair regulators contains more taxpayers than those overseen by corrupt regulators. This means that even with the same value of $m$, the total regulatory fees for fair regulators will exceed those of corrupt regulators. As $m$ increases, this disparity widens further. Thus, even if the bribes received by corrupt regulators exceed their monitoring cost $\gamma$, the payoff of fair regulators remains higher than that of corrupt regulators. 

Therefore, we conclude that improving the compensation of government officials helps to reduce corruption, thereby enhancing citizens' tax compliance. When officials' legitimate income is insufficient to meet their basic living needs, the likelihood of engaging in corrupt behavior significantly increases. The statistics show that in economically developed regions, the incidence of government corruption is generally lower than in undeveloped areas, and citizens' tax compliance is higher, forming a virtuous cycle of ``adequate compensation - clean governance - tax compliance''. Our simulation results are also consistent with this observation.

\subsection{The effect of monitoring cost $\gamma$ on $f_C$ and $f_F$}

As we argued, there is no free lunch, hence corrupt regulators should pay the price to be hidden. In the following, we reveal how the related monitoring cost $\gamma$ impact on $f_C$ and $f_F$. The resulting curves of tax compliance rate $f_C$ and the proportion of fair regulators $f_F$ against the synergy factor $r$ are shown in Fig.~\ref{fig6} at different $\gamma$ values. 

Our findings reveal that most curves initially start from a state of pure evasion and pure corruption when $r$ is small, followed by a rapid increase in both $f_C$ and $f_F$ values as $r$ grows, eventually reaching the state of pure compliance and pure fairness. The increase in monitoring cost $\gamma$ significantly reduces the extinction thresholds for evaders and corrupt regulators. Notably, when $\gamma=0$, the emergence threshold for taxpayers in the game layer is $r=2.65$, while the regulatory layer remains in a state of pure corruption. This occurs because corrupt regulators do not incur costs for their corrupt behaviors, allowing them to get higher payoffs than fair regulators. This payoff advantage triggers a wave of strategy imitation in the regulatory layer, accelerating the spread of corrupt behavior. In the game layer, even under the supervision of completely corrupt regulators, taxpayers can cluster together to achieve higher payoffs than neighboring evaders, thereby maintaining tax-compliant behavior. This suggests that even a fully corrupt intervention system can improve social efficiency when $r$ is high enough. However, taxpayers are not always able to resist the incursions of evaders, which we discuss in detail later. 

From Fig.~\ref{fig6}, we can conclude that increasing the additional cost for corrupt officials can serve as a deterrent, thereby reducing the occurrence of corrupt practices. In reality, many countries have already combated corruption by implementing stricter laws and harsher punitive measures.

\subsection{The tax compliance and fairness landscapes on $m-\gamma$ parameter space}

To explore the combined effects of the regulatory fee $m$ and the monitoring cost $\gamma$ on the evolution more generally, we summarize the key quantities on the mentioned parameter plane shown in Fig.~\ref{fig7}. Panels~(a) and (b) present the values of $f_C$ and $f_F$ at parameter conditions $\alpha=1.4$ and $\beta=0.3$, while panels~(c) and (d) display the corresponding results for $\alpha=1.2$ and $\beta=0.4$. 

As a general observation, both $f_C$ and $f_F$ grow when we increase the regulatory fee or the corruption cost. For each specified $m$, as $\gamma$ increases, both $f_C$ and $f_F$ rise, which is consistent with our previous analysis, and similar agreement is found when $\gamma$ is fixed while $m$ varies. Meanwhile, the fraction of taxpayers and the level of fairness show similar trends on the heatmap. Specifically, when the value of $\gamma$ is very small, the proportion of fair regulators is 0 in Fig.~\ref{fig7:b} and in Fig.~\ref{fig7:d}. The proportion of taxpayers is also 0 in Fig.~\ref{fig7:a}, while it reaches approximately 0.5 in Fig.~\ref{fig7:c}. This occurs because in the top row, the penalty $\alpha=1.4$ and bribery ratio $\beta=0.3$, meaning evaders under corrupt regulators only need to pay a bribe of $\alpha\beta=0.42$. In contrast, in the bottom row where corrupt regulators dominate the regulatory layer, evaders must pay a higher bribe of $\alpha\beta=0.48$ (derived from $\alpha=1.2$ and $\beta=0.4$). For some evaders, the excessively high bribery cost reduces their payoffs below those of neighboring taxpayers, leading them to switch to tax compliance in the next round. 

The results indicate that increasing government official's salary and strengthening the fight against corruption both contribute to the establishment of a clean government, thereby enhancing the willingness of citizen's tax compliance.

\subsection{Coevolutionary pattern formation of tax compliance and regulator fairness}

\begin{figure*}[t]
    \centering
    \begin{subfigure}[b]{0.24\textwidth}
        \centering
        \includegraphics[width=\textwidth]{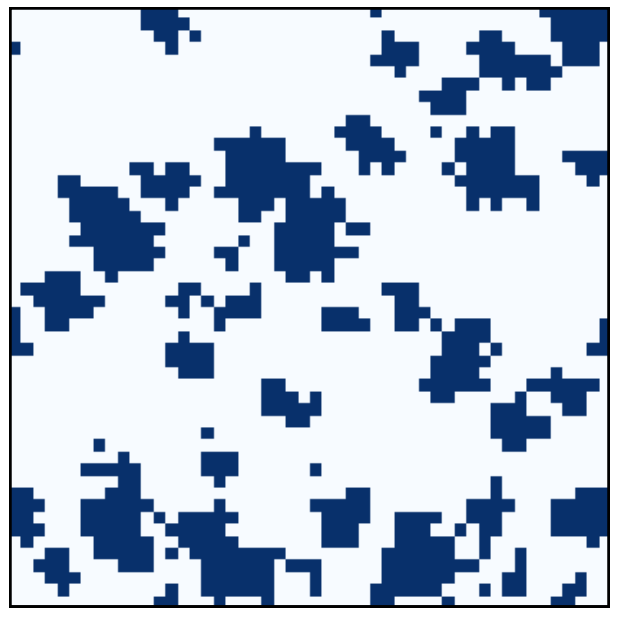}
        \caption{$t=10$}
        \label{fig8:a}
    \end{subfigure}
    \hfill
    \begin{subfigure}[b]{0.24\textwidth}
        \centering
        \includegraphics[width=\textwidth]{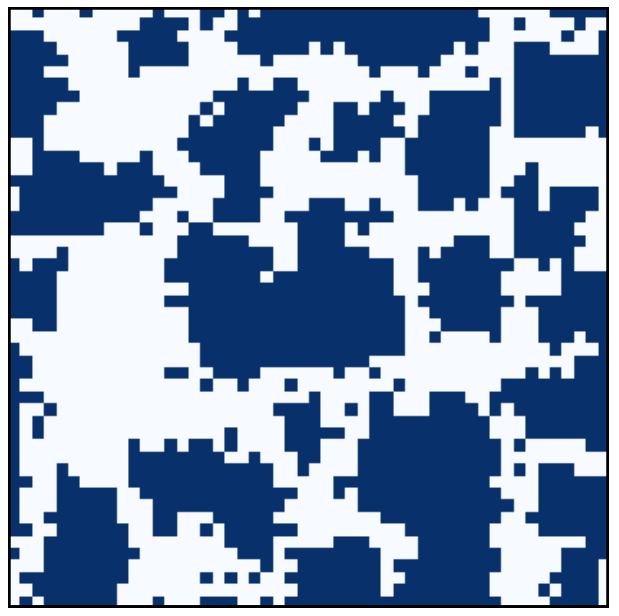}
        \caption{$t=200$}
        \label{fig8:b}
    \end{subfigure}
    \hfill
    \begin{subfigure}[b]{0.24\textwidth}
        \centering
        \includegraphics[width=\textwidth]{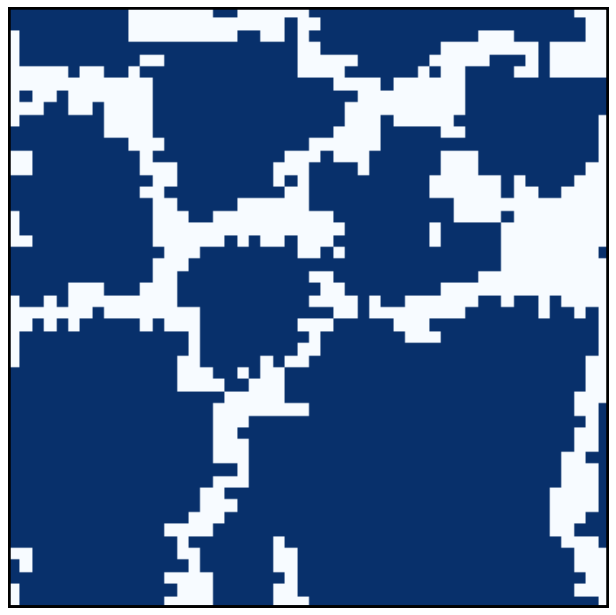}
        \caption{$t=1000$}
        \label{fig8:c}
    \end{subfigure}
    \hfill
    \begin{subfigure}[b]{0.24\textwidth}
        \centering
        \includegraphics[width=\textwidth]{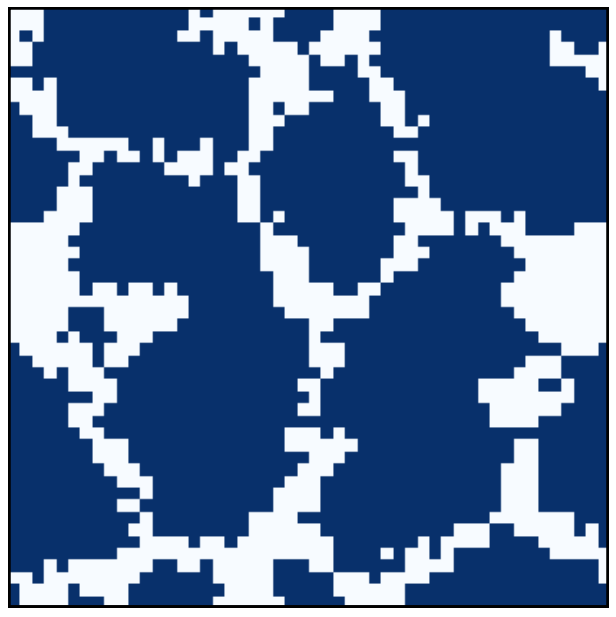}
        \caption{$t=3000$}
        \label{fig8:d}
    \end{subfigure}

    \vspace{\baselineskip}

    \begin{subfigure}[b]{0.24\textwidth}
        \centering
        \includegraphics[width=\textwidth]{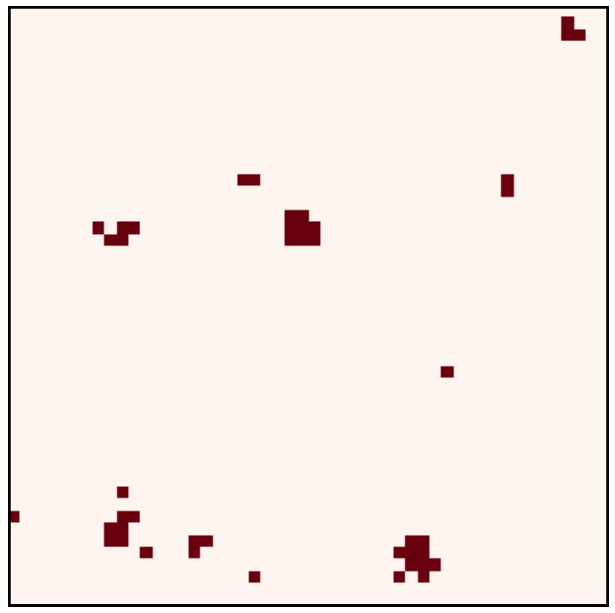}
        \caption{$t=10$}
        \label{fig8:e}
    \end{subfigure}
    \hfill
    \begin{subfigure}[b]{0.24\textwidth}
        \centering
        \includegraphics[width=\textwidth]{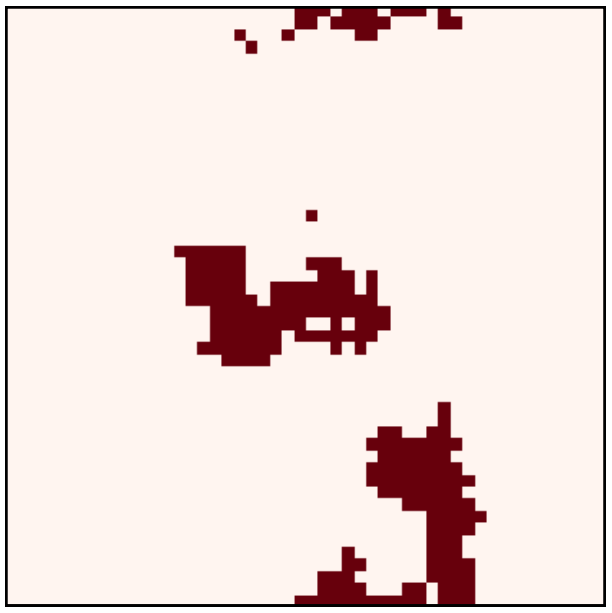}
        \caption{$t=200$}
        \label{fig8:f}
    \end{subfigure}
    \hfill
    \begin{subfigure}[b]{0.24\textwidth}
        \centering
        \includegraphics[width=\textwidth]{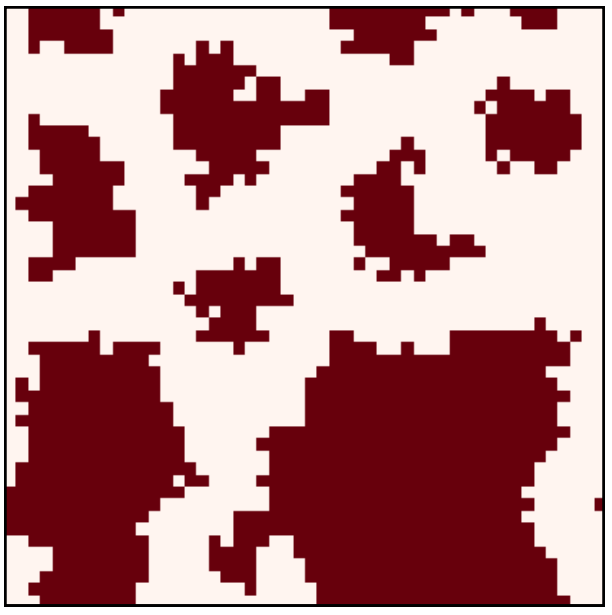}
        \caption{$t=1000$}
        \label{fig8:g}
    \end{subfigure}
    \hfill
    \begin{subfigure}[b]{0.24\textwidth}
        \centering
        \includegraphics[width=\textwidth]{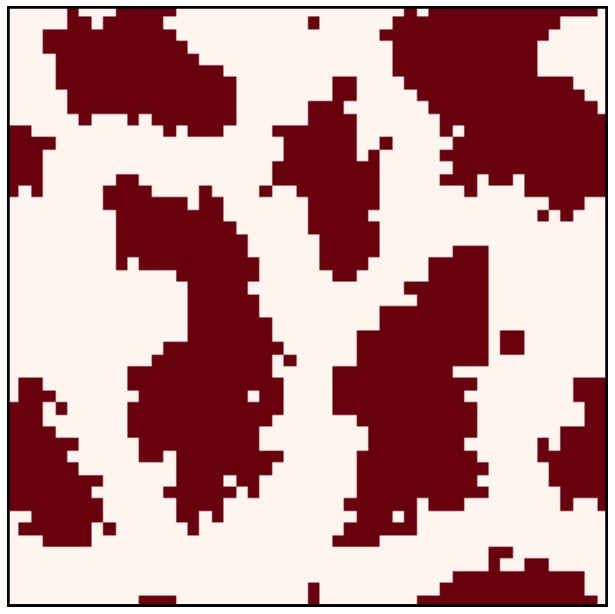}
        \caption{$t=3000$}
        \label{fig8:h}
    \end{subfigure}
    \caption{\textbf{Coevolution of tax compliance and regulator fairness.} Starting from a random initial state, panels~(a)-(d) show the evolution of strategy distributions in the game layer obtained at $t\in\{10,200,1000,3000\}$ respectively. Here, blue (\textit{resp.} white) cells represent taxpayer (\textit{resp.} evader). In the bottom row, panels~(e)-(h) illustrate how the patterns of competing regulators evolve in the regulatory layer obtained at the same steps specified above. Here, red (\textit{resp.} linen) color represents fair (\textit{resp.} corrupt) regulators. Other parameters are $r = 2.0$, $\alpha = 1.2$, $\beta = 0.4$, $m = 0.1$ and $\gamma = 0.1$. The comparison reveals a strong correlation, tax compliance and regulator fairness evolve hand in hand by supporting each other.}
    \label{fig8}
\end{figure*}

To understand the interdependence of tax compliance and regulator fairness more deeply, we study the pattern formation on both layers simultaneously. 

In Fig.~\ref{fig8}, we present how the strategy distribution of individuals evolves in time, both in the game layer and in the regulatory layer. We record the actual distributions obtained after 10, 200, 1000, and 3000 MC steps. At the very beginning, shown in panel~(a) and in panel~(e), the game layer's evaders and regulatory layer's corrupt regulators dominate the system due to the random initial configuration creating optimal conditions for selfish behavior to proliferate. More precisely, evaders can obtain higher payoffs from the public pool than the penalties or bribes they need to pay, which incentivizes taxpayers to adopt their evasion strategy. As the number of evaders increases, corrupt regulators collect more bribes, resulting in significantly higher payoffs compared with fair regulators. Consequently, the regulatory layer becomes predominantly occupied by corrupt regulators, with only a minimal presence of fair regulators remaining. Later, the remaining taxpayers in the network cluster together to resist the incursions of evaders, gradually highlighting the advantages of tax compliance. The payoffs of evaders become insufficient to cover fines or bribes, leading to a reduction in their numbers. At this point, in the regulatory layer, the earnings of corrupt regulators will decrease as the number of evaders in the game layer declines. When a corrupt regulator supervises a PGG group consisting entirely of taxpayers, they not only lose all bribery income but must also bear the additional cost of their corrupt behavior, making the transition to a fair strategy their optimal choice. Furthermore, we find that in areas with a dense distribution of fair regulators, the corresponding game layer also exhibits a significant clustering of taxpayers, which becomes more pronounced after evolutionary stabilization. In this way, the mentioned actors can support each other across the layers and tax compliance and fairness can evolve hand in hand. However, the area of clusters formed by fair regulators is always smaller than that of taxpayers. This is because, in mixed PGGs of taxpayers and evaders, evaders exert a strong temptation on regulators through bribery. At the same time, regulators are more likely to learn from their surrounding corrupt regulators (who oversee PGGs composed entirely of evaders) due to their high payoffs. This dual mechanism ultimately leads to a noticeable tendency toward corruption in the strategy choices of regulators in mixed PGGs. It is worth noting that the information flow between the layers via bribe and punishment makes a correlated strategy change, hence a coordinated evolution occurs. Similar simultaneous growths were reported in~\citep{szolnoki2013}, highlighting the robustness of our present findings. 

In real-world governance, the government should prioritize supervising regions with concentrated tax evaders and corrupt regulators, as these constitute high-risk zones particularly prone to corruption and requiring intensified oversight. Such areas exhibit two characteristics: first, regulators face strong bribery incentives, as evaders are highly motivated to bribe their way out of penalties; second, regulatory strategy selection demonstrates a ``bad money drives out good'' effect, where fair regulators are increasingly assimilated through imitation of corrupt peers. Consequently, governments must move beyond simplistic ``punishment-as-governance'' approaches and instead establish an integrated framework combining penalties, incentives, and constraints. This systemic strategy must also account for the networked diffusion patterns of corrupt behavior, which propagate through both economic transactions and social learning. 

\section{Conclusions and Discussion}\label{sec4}

Tax compliance behavior, as a typical social cooperation phenomenon, has long attracted the attention of economics, sociology, and complex systems science regarding its evolutionary mechanisms. This study constructs an interdependent two-layer network model to explore the co-evolutionary mechanisms of tax compliance and government regulation within the spatial public goods game. The core innovation of the model lies in considering taxpayers' strategic interactions and regulators' behavioral heterogeneity. In the game layer, taxpayers decide whether to fulfill their tax obligations through a PGG. In the regulatory layer, government officials choose between impartial enforcement or bribe acceptance, creating a dynamic regulation-feedback mechanism. Moreover, for corrupt government officials, they also need to bear the additional cost of their misconduct when accepting bribes, which makes the model more realistic. Our simulation results demonstrate that the penalty $\alpha$ and bribery ratio $\beta$ jointly influence the degree of tax compliance and fair regulator intensity. As the value of $\alpha$ increases, the game layer exhibits a rapid phase transition from pure evasion to pure compliance, while the regulatory layer simultaneously undergoes a transformation from pure corruption to pure fairness. The influence of $\beta$ on tax compliance rate and fair regulators density is nonlinear, arising from its dual effects on the payoffs of both evaders and corrupt regulators. In the low $\beta$ range, a decrease in the bribery ratio reduces the violation costs for evaders but diminishes the additional payoffs corrupt regulators obtain through bribery. Conversely, in the high $\beta$ range, as bribery expenditures increase, the evader group gradually shrinks, leading to a reduction in the illegal incomes for corrupt regulators. Ultimately, $\beta$ exhibits an initial suppressive and then a facilitative effect on the evolution of tax compliance, providing a new perspective to understand the complex roles of bribery and corruption in tax governance. The regulatory cost $m$ and monitoring cost $\gamma$ also influence the emergence of fair regulators. When $m$ is low, the relative disadvantage in payoffs for fair regulators leads to the spread of corrupt behavior in the regulatory layer. Additionally, the value of $\gamma$ determines the risk-reward ratio for regulators' behaviors. Inadequate corruption costs significantly lower the threshold for corrupt behavior, making it difficult to maintain fairness. 

We also analyze the snapshots of the game and regulatory layer, the spatial structure of fair regulators shows the same clustering patterns as taxpayers do, which defends the invasion of evaders and corrupt regulators. The findings provide several insights for tax policy design. First, strengthening penalty enforcement can deter evasion only when institutional fairness and public trust are maintained. Second, improving regulators’ compensation must be accompanied by transparent oversight to ensure integrity. Third, our results highlight the importance of balancing deterrence and incentive mechanisms in tax governance.

While the present model provides valuable theoretical insights, it is subject to certain limitations. The punishment fines and bribes can be a nonlinear function. In reality, the compensation structure for government officials is complex and not as simplified as presented in this paper, and the monitoring costs faced by each corrupt official are not uniform. Additionally, for simplification, we assumed that the game layer and regulatory layer are identical. But it is evident that the number of officials and citizens cannot be the same in reality. In higher-order networks, the impacts of fair and corrupt regulators may also differ substantially from our current results. Moreover, the simulations assume homogeneous agents, uniform monitoring costs, and idealized network structures, which may differ from real-world administrative systems. Future studies could extend this framework by incorporating heterogeneous behavioral traits, multi-level governance, and empirical calibration using survey or administrative data. Such extensions would deepen understanding of how institutional design and social norms co-evolve to sustain compliance.

\bmhead{Funding Declaration}

This work is supported by the Chongqing Social Science Planning Project under Grant No. 2025NDQN41, Natural Science Foundation of Chongqing under Grant No. CSTB2025YITP-QCRCX0007, and National Research, Development and Innovation Office (NKFIH) under Grant No. K142948.

\bmhead{Competing interests}
The authors declare no competing interests. Minyu Feng was a member of the Editorial Board of this journal at the time of acceptance for publication. The manuscript was assessed in line with the journal’s standard editorial processes, including its policy on competing interests.

\bmhead{Data availability}
All codes developed in this study have been deposited into a publicly available GitHub repository at https://github.com/ting-2727/taxation.git.

\bmhead{Ethical approval}
Ethical approval was not required as the study did not involve human participants.

\bmhead{Informed consent}
Informed consent was not required as the study did not involve human participants.

\bmhead{Author contributions}
Qin Li: Idea formulation, model conception, and theoretical framework development;  
Ting Ling: Simulation, parameter tuning, data analysis, drafting of the original manuscript, and partial contribution to methodological design; 
Minyu Feng: Paper supervision, conceptual guidance, manuscript revision, editing, final approval, and funding acquisition; 
Attila Szolnoki: Methodological refinement, editing, and funding acquisition.

%\bibliographystyle{plainnat} % 自动生成作者-年份格式

\bibliography{sn-bibliography}% common bib file

\end{document}